\documentclass[12pt]{article}

\usepackage[T1]{fontenc}
\usepackage[utf8]{inputenc}
\usepackage[english]{babel}
\usepackage{amsmath,amssymb,amsfonts,amsthm}        
\usepackage{mathtools,mathrsfs}                     
\usepackage{thmtools,tensor,bbm,bm}                 
\usepackage[makeroom]{cancel}                       
\usepackage{geometry,fancyhdr}                      
\usepackage{graphicx,caption,subcaption}            
\usepackage[dvipsnames,svgnames,x11names]{xcolor}   
\usepackage[pdfencoding=auto, psdextra]{hyperref}   
\usepackage{cite}                                   
\usepackage[capitalize]{cleveref}                   
\usepackage{etoolbox,tcolorbox}                     

\newcommand\myshade{90}
\colorlet{mycitecolor}{JungleGreen}
\colorlet{mylinkcolor}{purple}
\colorlet{myurlcolor}{MediumSeaGreen}
\hypersetup{
    colorlinks = true,
    citecolor  = mycitecolor!\myshade!black,
    linkcolor  = mylinkcolor!\myshade!black,
    urlcolor   = myurlcolor!\myshade!black
}

\renewenvironment{abstract}
    {\begin{center}\bfseries Abstract.\end{center}\vspace{-30pt}
    \quotation\small\noindent\rule{\linewidth}{.5pt}\par\medskip}
    {\par\noindent\rule{\linewidth}{.5pt}\endquotation}
    \AtBeginEnvironment{abstract}{\def\quotation{\list{}{\rightmargin\leftmargin}\item\relax}\def\endquotation{\endlist}}

\newtheoremstyle{definitionstyle}
  {\topsep}
  {\topsep}
  {}
  {0pt}
  {\bfseries}
  {.}
  { }
  {\thmname{#1}\thmnumber{ #2}\thmnote{: #3}}

\newtheoremstyle{examplestyle}
  {\topsep}
  {\topsep}
  {}
  {0pt}
  {\bfseries}
  {.}
  { }
  {\thmname{#1}\thmnumber{ #2}\thmnote{: #3}}

\theoremstyle{definitionstyle}
\newtheorem{definitioninner}{Definition}[section]
\newenvironment{definition}[1][]{
  \begin{definitioninner}[#1]
  \hspace{0.1em}}
  {\hfill$\square$
  \end{definitioninner}}

\theoremstyle{examplestyle}
\newtheorem{exampleinner}{Example}[section]
\newenvironment{example}[1][]{
  \begin{exampleinner}[#1]
  \hspace{0.1em}}
  {\hfill$\blacksquare$
  \end{exampleinner}}

\newcommand{\C}{\mathbb{C}}

\newcommand{\R}{\mathbb{R}}
\newcommand{\Tr}[1]{\:{\rm Tr}\,#1}
\newcommand{\dd}{{\mathrm d}}
\newcommand{\del}{\partial}

\newcommand{\eqn}[1]{(\ref{#1})}

\DeclareMathOperator{\tr}{Tr}
\DeclareMathOperator{\DD}{D}

\title{\vspace{-2cm}
    \textbf{Introduction to noncommutative field and gauge theory}
    \vspace{0pt}}
\date{
    \vspace{-20pt}}
\author{
	Patrizia Vitale,$^{(a)}$ Martina Adamo,$^{(b)}$ Roukaya Dekhil,$^{(c)}$ \\
    Diego Fernández-Silvestre$^{(d)}$
	\vspace{5pt}\\
    \small $^{(a)}$Dipartimento di Fisica “E. Pancini”, Università di Napoli Federico II\\
    \small and INFN, Sez.~di Napoli, Italy \\
	\small $^{(b)}$Departamento de F\'isica, Universidad de Burgos, 09001 Burgos, Spain \\
    \small $^{(c)}$Arnold Sommerfeld Center for Theoretical Physics, Ludwig-Maximilians-Universit\"at München,\\ \small Theresienstrasse 37, 80333 M\"unchen, Germany \\
    \small $^{(d)}$Departamento de Matem\'aticas y Computaci\'on, Universidad de Burgos, 09001 Burgos, Spain
    \vspace{10pt}}

\begin{document}

\maketitle
\begin{abstract}
These are lecture  notes for an introductory course on  noncommutative field and gauge theory.  We begin by reviewing  quantum mechanics as the prototypical  noncommutative  theory, as well as   the geometrical language of  standard gauge theory. Then, we review a specific approach to noncommutative field and gauge theory, which relies on the introduction of a derivations-based differential calculus. We focus on the cases of  constant and linear noncommutativity, \textit{e.g.}, the Moyal spacetime  and the so-called $\R^3_\lambda$, respectively. In particular, we review  the $g\varphi^4$ scalar field theory and the $U(1)$ gauge theory on such noncommutative spaces. Finally, we discuss  noncommutative spacetime symmetries from both the observer and particle point of view. In this context, the twist approach is reviewed and the $\lambda$-Minkowski $g\varphi^4$ model is discussed.
\end{abstract}
\tableofcontents


\section{Introduction}
These lecture notes have been prepared on the basis of  a short introductory course on noncommutative field and gauge theory delivered at the Second Training School
of COST Action CA18108 ``Quantum gravity phenomenology in the multi-messenger approach''. They are not meant to be exhaustive, as they reflect the personal journey of the speaker through the development of the subject. Important approaches to the subject, such as the {\it spectral action} formulation of the standard model started by  Connes and Chamseddine \cite{Connes:2006qj}, as well as pioneering  contributions by Madore,  Wess, and  collaborators \cite{Madore:2000en}, or the matrix-model approach to gauge theory and emergent gravity \cite{Steinacker:2003sd}, are missing, and we refer the interested reader to the cited literature.

The exploration of noncommutative spacetime, matter, and gauge fields is driven by a range of motivations, each offering insights into different branches of theoretical physics as well as mathematics. One key motivation is its potential to be a signature of quantum gravity \cite{Bronstein:2012zz}. In this context, a classical argument is the \textit{gedanken} experiment described in \cite{Doplicher:1994tu}, which points to the breakdown of the Riemannian structure of spacetime at scales where the predictions of both quantum mechanics and general relativity are brought together. Another classical motivation, which goes back to the foundations of quantum mechanics \cite{Heisenberg:1938xla, Psnyder}, is the appealing idea that spacetime noncommutativity implies a minimal area, namely the disappearance of point-like objects. This would solve the problem of the ultraviolet regime of quantum field theory (QFT), addressing infinite energy issues that arise at extremely small scales. This regularization, as we shall briefly see, introduces, however, new complexities in quantum field theory.

More recently, different theoretical approaches to quantum gravity have suggested that spacetime may not possess the smooth usual structure of a differentiable manifold. For instance, in background-independent approaches such as loop- and  spin foam-quantum gravity or group field theory \cite{Freidel:2005qe, Oriti:2006ar}, the quantum operators associated with the geometric quantities such as the area and volume exhibit discrete spectra, indicating a departure from the smooth manifold structure (at the Planck scale)  usually assumed \cite{Ashtekar:1996eg, Ashtekar:1997fb}. Last but not least,  the appearance of noncommutative geometry in string theory with a nonzero B-field is an important result which has produced an enormous boost to the research activity in the field: in a  famous paper in 1999 \cite{Seiberg:1999vs} Seiberg identifies a limit in which the string dynamics is described by a minimally coupled (supersymmetric) gauge theory on a noncommutative space. 

The paper is organized as follows. In Section \ref{DFR argument} we shortly review the  Doplicher--Fredenhagen--Robert (DFR) argument that justifies the introduction of a noncommutative spacetime. Then, to illustrate in a familiar situation what a noncommutative geometry is, we review in Section \ref{phasespace}  the Weyl--Wigner--Moyal formulation of quantum mechanics, stressing the quantum nature of phase space, which is not a smooth manifold anymore but indeed a noncommutative geometry.  

Before abandoning the classical picture of spacetime, we summarize in Section \ref{sec:cgft}  the basic geometric ingredients of commutative gauge theory and sketch their algebraic counterpart, which, in Section \ref{sec:ncgft}, will serve as a basis for the introduction of their noncommutative generalization.  
In Section \ref{sec:differential calculus} we extend the algebraic description of  the differential calculus to the noncommutative setting. Hence, two instances of noncommutative spaces are considered, \textit{e.g.}, the Moyal spacetime and Lie-algebra type noncommutative spaces.  

We will devote Section \ref{ScalarF_on_moyal} and Section \ref{DSec2} to studying the dynamics of scalar field theory and gauge fields on the Moyal space. As for the gauge theory, we will only consider electrodynamics, although the approach remains valid for non-abelian Yang--Mills theories.   In Section \ref{DSec1} we move to the next simplest case and consider linear-type noncommutativity, leading to the construction of the  $\R^3_\lambda$ algebra, together with its differential calculus, and we analyze an interacting scalar field theory on such space. Proceeding by increasing the complexity, we thus consider, in \ref{DSec4}, the $U(1)$ gauge theory on such space.  
 
To conclude, we discuss the spacetime symmetries in the noncommutative context. We distinguish between the passive and active points of view and show in Section \ref{observersymm} that the Moyal product is covariant under passive or observer-dependent Poincar\'e transformations, hence it is possible to construct gauge-invariant actions. This gives the cue to discuss active or particle-dependent transformations within the twist approach to noncommutative models, summarized in Section \ref{DSec3} and applied to the so-called $\lambda$-Minkowski spacetime in Section \ref{DSec5}.  

Concluding remarks and two appendices complete the manuscript.

\section{The Doplicher--Fredenhagen--Robert argument} 
\label{DFR argument}

It is widely believed that our understanding of spacetime as a manifold, which resembles the flat Minkowski spacetime on a local scale, breaks down at very short distances, such as the Planck length $\lambda_P=\sqrt{\hbar \, G/c^3}\simeq 10^{-35}$ m. At such tiny scales, the accuracy of localizing spacetime events faces some limitations, particularly when incorporating gravitational effects into a quantum theory.
In their article \cite{Doplicher:1994tu}, the authors showed that there are uncertainty relations among the coordinates of spacetime events. Similar arguments were already discussed by Bronstein \cite{Bronstein:2012zz}. Their derivation stems from Heisenberg's principle and Einstein's classical theory of gravity.  The attempt to achieve precise localization encounters gravitational collapse, making spacetime below the Planck scale devoid of any operational meaning. Building upon this observation, the authors of \cite{Doplicher:1994tu} develop spacetime uncertainty relations. According to their proposal, spacetime possesses an inherent quantum structure that naturally gives rise to these relations. Hence, the operational meaning problem at small scales is a built-in feature of the model: spacetime as a noncommutative manifold.

\paragraph{Spacetime uncertainties.} According to Heisenberg's uncertainty principle, the precise measurement of spacetime coordinates with an accuracy of $a$ results in an uncertainty in the associated momentum of the order of $1/a$ (assuming natural units $\hbar = c = G = 1$, unless otherwise specified). Neglecting rest masses, this measurement process leads to the transfer of energy of the order $1/a$, which becomes concentrated within the localization region at a specific time. As a consequence, an energy-momentum tensor $T_{\mu\nu}$ emerges, giving rise to a gravitational field. In principle, this field should be determined by solving Einstein's equations for the Minkowski metric $\eta_{\mu\nu}$,
\begin{equation}
    R_{\mu\nu} - \tfrac{1}{2}R \, \eta_{\mu\nu} = 8\pi \, T_{\mu\nu} \, .
\end{equation}
Therefore, as the uncertainty $\Delta x_\mu$ in coordinate measurements decreases, the gravitational field generated by the measurement intensifies. In the attempt to localize the particle $\Delta x_\mu \to 0$  with high precision the gravitational field becomes so strong that it prevents light or other signals from escaping within the region under consideration. This means that the act of measurement itself loses meaning, and so does the notion of localization. Hence, in order to ensure that no black hole is generated during the measurement process, the localization must be constrained. The values of $\Delta x_\mu$ will be subject to certain restrictions, preventing them from being simultaneously arbitrarily small,
\begin{equation}
    \Delta x_\mu \Delta x_\nu \geq \lambda_P^2 \, .
\end{equation}

\paragraph{Quantum (noncommutative) spacetime.} In \cite{Doplicher:1994tu} it was suggested that the noncommutativity of spacetime coordinates could provide an explanation for spacetime uncertainty relations, namely
\begin{equation}
    [ x_\mu, x_\nu ] \neq 0 \, .
\end{equation}
The coordinates $x_\mu$ are therefore operators with non-trivial commutation relations, describing a \textit{quantum} (or \textit{noncommutative}) \textit{spacetime}. Therefore, the observables, that were smooth functions on classical spacetime, are now represented as operators that generally do not commute. Similarly, the states, which were represented as points in classical spacetime (or, equivalently, as evaluation maps on the space of classical observables)\footnote{An evaluation map (associated to a classical state $w$) on the space of classical observables, is a map that at each smooth function on the classical spacetime, $f\in\mathcal{F}(M)$ (a classical observable), associates the value of $f$ at  $w\in M$,
\begin{equation}
    \delta_w\, :\, f\in\mathcal{F}(M)\to f(w)\in \mathbb{C} \,.
\end{equation}
}, are now described by quantum evaluation maps (\textit{i.e.}, linear functionals on the space of quantum observables).

\section{Phase-space formulation of quantum mechanics} \label{phasespace}

Quantum mechanics can be seen as a noncommutative geometry in phase space. Unlike classical mechanics, where we have a smooth phase space, in quantum mechanics the phase space has a minimal area (represented in units of $\hbar$) which arises from the uncertainty principle. The coordinate functions $q$ and $p$ no longer commute, leading to a departure from the notion of a differentiable manifold. Moreover, in quantum mechanics classical observables are replaced by Hermitian operators, and the classical states, which correspond to points in the classical phase space, are now described as vectors in a Hilbert space. In the following, we summarize the Weyl--Wigner--Moyal approach (WWM) to quantum mechanics \cite{Weyl, Wigner, Gronewold, Moyal}, that allows a convenient parallel with quantum spacetime as a noncommutative geometry. It is often referred to as {\it phase-space formulation} of quantum mechanics because operators and quantum density states are replaced by noncommuting functions on classical phase space (namely, operator symbols on the cotangent bundle $T^*M$), 
using quantizer and dequantizer operators \cite{Manko:2004syv}. Through dequantization, we can achieve a completely faithful description of quantum mechanics by replacing quantum observables with a noncommutative algebra of functions with a $\star$-product (the Moyal product or its siblings) and quantum density states by quasi-probability distributions (the Wigner functions).

\paragraph{Operator symbols.} In classical statistical mechanics, finding the average value of an observable, like particle energy, requires averaging the corresponding function $E(q,p)$ using the probability density $f(q, p)$,
\begin{equation}
    \langle E \rangle = \int \! \dd q \dd p \, E(q,p) f(q,p) \, .
\end{equation}
In quantum mechanics, observables are represented by Hermitian operators. Similarly to statistical mechanics, to calculate the expectation value of an operator $\hat{A}$ in a given quantum state $\hat\rho$, we can similarly associate it with a function $f_{\hat{A}}(q,p)$ defined on the phase space. This function is known as the \textit{symbol of the operator} $\hat{A}$, and it enables us to calculate the mean value of the operator (in a state described by the quasi-probability distribution $W(q, p)$) as follows:
\begin{equation}
    \langle \hat{A} \rangle = \int \! \dd q \dd p \, f_{\hat{A}}(q,p) W(q,p) \, .
\end{equation}
In the WWM approach, the quasi-probability distribution $W(q,p)$ is the Wigner function associated with the density operator $\hat\rho$, whereas the operator symbol $f_{\hat{A}}$ is defined through the inverse Weyl map. The WWM and other quantization-dequatization schemes, corresponding to different operator orderings, are better understood within the procedure described below.

\paragraph{Quantizers and dequantizers.} Let us consider a family of Hermitian positive-definite operators on the phase space, $\hat{U}(q,p)$, and an observable $\hat{A}$, labeled by the phase-space parameters $(q,p)$. We can construct the symbol of $\hat{A}$ as
\begin{equation} \label{dequantizer}
    \hat{A} \to f_{\hat{A}} (q,p)=\tr (\hat{A}\,\hat{U}(q,p)) \,.
\end{equation}
The operator $\hat{U}(q,p)$ is called \textit{dequantizer}\footnote{It is important to note that, as the dequantizer is simply required to be a Hermitian operator, the resulting symbol can exhibit negative values and this prevents its interpretation as measurement probability.} and the map defined in Eq.~\eqref{dequantizer} is linear,
\begin{equation}
    \begin{aligned}
        \hat{A}+\hat{B} &=\hat{C} \\
        \zeta  \hat{A} &= \hat{B}
    \end{aligned}
    \quad
    \begin{array}{c}
        \Leftrightarrow \\ \Leftrightarrow
    \end{array}
    \quad
    \begin{aligned}
        f_{\hat{A}}(q,p)+f_{\hat{B}}(q,p)& =f_{\hat{C}}(q,p) \,, \\
        \zeta f_{\hat{A}}(q,p) &= f_{\hat{B}} (q,p) \,.
    \end{aligned}
\end{equation}
Moreover, if the map is invertible so as to have a one-to-one correspondence between the operator and its symbol, then the latter contains all the information about the operator, which can be therefore reconstructed through another family of Hermitian operators, $\hat{D}(q,p)$,
\begin{equation} \label{quantizer}
    f_{\hat{A}} (q,p) \to \hat{A}=\int\! \dd q \dd p \, f_{\hat{A}}(q,p) \hat{D}(q,p) \,.
\end{equation}
This is the inverse map of \eqref{dequantizer} and  $\hat{D}(q,p)$ is called \textit{quantizer}, with the compatibility condition
\begin{equation}
\Tr \hat U(q,p)\hat D(q', p') + \delta(q-q')\delta(p-p')\,.
\end{equation}

\paragraph{The Weyl--Stratonovich operator.} In the WWM approach, or symmetric ordering, quantizer, and dequantizer operators are essentially the same, meaning that $\hat{U}(q,p)=2\pi \hat{D}(q,p)=\hat{\Omega}(q,p)$, for
\begin{equation}
\hat{\Omega}(q,p)=\int \! \dd\eta \dd\xi \, e^{i(\eta \hat{Q}+\xi \hat{P})} e^{-i(\eta q+\xi p)}\,,
\end{equation}
which is the so called \textit{Weyl--Stratonovich operator}. Here $\hat{Q}$ and $\hat{P}$ are the position and momentum operators associated with the phase-space coordinates $q$ and $p$, respectively. For this reason, this quantization scheme is also called self-dual. It is important to note that for other ordering prescriptions the quantizer and dequantizer are different (see for example \cite{Manko:2004syv, Lizzi:2014pwa} for details).

\paragraph{Quantum states.} Quantum states, both pure and mixed,  are conveniently described in terms of density operators.  The corresponding symbol is in general a  quasi-probability distribution defined as
\begin{equation}
 \hat{\rho} \to W_{\hat{\rho}}(q,p) = \tr (\hat{\rho}\,\hat{D}(q,p)) \,.   
\end{equation}
whereas the inverse map is obtained through the operator $\hat U$.
Notice the inverted role played by the operators $\hat U$ and $\hat D$ in relation to the states, as compared with \eqn{dequantizer}, \eqn{quantizer}. It is only in the WWM scheme that states and observables are quantized/dequantized by the same operators, while in general states and observables, being dual objects, are dealt with in a dual approach \cite{Manko:2006icr}. The symbol of the density operator through the Weyl--Stratonovich operator is the famous Wigner quasi-probability distribution, but other well-known quasi-distributions corresponding to normal and antinormal orderings are the Husimi and Glauber--Sudarshan functions \cite{Manko:2004syv}.

\paragraph{Star-product.} The set of operator symbols inherits a noncommutative product, hence an algebra structure, which is given by the operator product
\begin{equation}
    \hat{A} \hat{B} \to (f_{\hat{A}} \star f_{\hat{B}}) (q,p) = \tr (\hat{A}\hat{B}\,\hat{U}(q,p))\,,
\end{equation}
known as \textit{star-product}. Associativity and noncommutativity descend from the operator product:
\begin{equation}
    \begin{array}{c}
        \text{associativity} \\
        \text{noncommutativity}
    \end{array}
    \quad
    \begin{aligned}
        (\hat{A}\hat{B}) \hat{C} &=\hat{A} (\hat{B} \hat{C}) \\
        \hat{A} \hat{B} &\neq \hat{B}  \hat{A}
    \end{aligned}
    \quad
    \begin{array}{c}
        \Leftrightarrow \\ \Leftrightarrow
    \end{array}
    \quad
    \begin{aligned}
        (f_{\hat{A}}\star f_{\hat{B}}) \star f_{\hat{C}}& = f_{\hat{A}}\star (f_{\hat{B}} \star f_{\hat{C}}) \,, \\
        f_{\hat{A}}\star f_{\hat{B}} &\neq f_{\hat{B}}\star f_{\hat{A}} \,.
    \end{aligned}
\end{equation}
In particular, this yields a non-trivial commutator between phase-space coordinates,
\begin{equation}\label{qpnc}
    [q,p]_\star=q\star p - p \star q = i\hbar \,,
\end{equation}
where we restored  $\hbar$ for clarity. 

The phase-space formulation of quantum mechanics as a noncommutative geometry provides a comprehensive description of this theory, including its dynamics. This formulation yields a fully equivalent picture of quantum mechanics in terms of an algebra of noncommutative smooth functions over phase space, $(\mathcal{F}(T^* M), \star)$.

\section{Commutative gauge theory} \label{sec:cgft}

Gauge theories are field theories describing fundamental interactions. The so-called gauge groups are infinite-dimensional groups modeled on unitary Lie groups, abelian or non-abelian, depending on the interaction they describe. Since the underlying spacetime is commutative (generally Minkowski), we shall refer to them as commutative (abelian or non-abelian) gauge theories as opposed to noncommutative gauge theory (abelian or non-abelian) which has a noncommutative underlying spacetime.  
Transformations that leave a physical system unchanged under their action are associated with symmetries. Gauge invariance is a local symmetry, meaning that the group parameters depend on spacetime coordinates. Therefore, the gauge group can be thought of as constructed by attaching a copy of a finite-dimensional Lie group to each point in spacetime (this is loosely speaking a principal fiber bundle over spacetime, with fibers being modeled on a finite-dimensional Lie group called the \textit{structure group}). Thus, gauge transformations may be defined as maps from the spacetime manifold $M$ to the structure group $G$ \cite{hamilton2017mathematical,rubakov2009classical,sundermeyer2014symmetries},
\begin{equation} 
\label{gaugetransf}
    \hat{G}=\{ g : \, x\in M \to g(x)\in G \}\,.
\end{equation}
Gauge theories describe the behavior of the fundamental interactions, such as the electromagnetic force, and the strong and weak nuclear forces. In the Standard Model picture, these fields are associated with bosons, which act as mediators for the corresponding fundamental interaction. Quantum electrodynamics (QED) is a $U(1)$-gauge theory describing the electromagnetic interaction. It has succeeded so far in explaining all phenomena involving electrically charged particles interacting through the exchange of photons, which are the bosons mediating this interaction. The weak interaction, on the other hand, is represented by the $SU(2)$-gauge theory called quantum flavor dynamics (QFD). It describes various phenomena including radioactive decay and nuclear fission/fusion, all mediated by the $W^{\pm}$ and $Z^0$ bosons. However, the weak interaction is better understood within the framework of the electroweak theory (EWT), which is a unified gauge theory that combines electromagnetic and weak interactions into a single model. EWT is based on a $SU(2)\times U(1)$ gauge symmetry. The strong interaction is the one responsible for binding quarks into hadrons and the formation of atomic nuclei. It is described by quantum chromodynamics (QCD), a $SU(3)$-gauge theory, and the associated gauge bosons are the gluons. Gauge theories with $U(N)$ structure group are known as \textit{Yang--Mills theories} \cite{manoukian2016quantum,scheck2018classical,marathe2010topics,sundermeyer2014symmetries}. Gravity, as described by General Relativity, is distinct from the other fundamental interactions. General Relativity explains gravity as the curvature of spacetime due to mass and energy. Although gravity is not traditionally formulated as a gauge theory, attempts have been made to incorporate it into this framework, \textit{e.g.}, as a $SO(1,3)$-gauge theory \cite{blagojević2013gauge}.

\subsection{Matter fields and group representations} \label{matterfield_commutative}

Matter fields represent particles that interact through the fundamental forces depicted by gauge theories. They possess the properties associated with the particles they describe, such as the charge, which determines in turn the type of interaction a particle experiences. For instance, particles with electric charges interact electromagnetically, while particles with weak charges engage in weak force interactions, and so forth. It is also possible for a particle to carry multiple charges, enabling it to couple with multiple gauge fields \cite{marathe2010topics,sundermeyer2014symmetries}. Mathematically, matter fields are described by vector fields on the spacetime manifold which carry a representation of the structure group, in particular, the defining representation. 
Therefore, the dimension of this representation determines also the dimension of the matter vector field \cite{rubakov2009classical,sundermeyer2014symmetries}.\\
The defining representations of $U(1)$, $SU(2)$, and $SU(3)$ are all complex and have dimensions $n=1,2,3$, respectively. This means that they are represented by complex matrices $M_{n\times n}$. Consequently, particles with electric charge $\phi^e$, weak charge $\phi^w$, or strong charge $\phi^s$ are represented by complex vector fields with one, two, and three components, respectively. This can be illustrated as follows:
\begin{equation}
    \begin{gathered}
        U(1) \, : \, \rho(g) \, \phi^e = M_{1\times1} \, \phi^e \,,  \qquad SU(2) \, : \, \rho(g) \, \phi^w = M_{2\times2} \left(\begin{array}{c} \phi^w_1 \\ \phi^w_2 \end{array}\right) \,, \\
        SU(3) \, : \, \rho(g) \, \phi^s = M_{3\times3} \left(\begin{array}{c} \phi^s_1 \\ \phi^s_2 \\ \phi^s_3 \end{array}\right) \,, \\
    \end{gathered}   
\end{equation}
where $g$ is an element of the gauge group associated with the considered interaction, and $\rho(g)$ denotes its representation on a matter field. Notice that, when a particle possesses multiple charges, it carries as many representations. For example, the electron is a complex one-dimensional vector field under $U(1)$, but it is a component of a doublet (the electron and its neutrino) under $SU(2)$ \cite{rubakov2009classical,sundermeyer2014symmetries}.

\subsection{Connection and curvature forms}

Let us consider a gauge theory with structure group $G$, defined on a four-dimensional manifold, which can have a Lorentzian or Euclidean signature. The fields of interest for a pure gauge theory, without matter fields, are the gauge potential $A$ and the field strength $F$. 
They are Lie-algebra-valued, meaning that they are endowed with internal/gauge indices in addition to the spacetime ones. Gauge indices are indicated by uppercase Latin letters $I, J, K,\dotsc =1,\dots, N$, where $N$ is the dimension of $G$, whereas for spacetime indices we use Greek letters $\mu, \nu, \rho, \dotsc=0,1,2,3$.
By denoting $\Omega^r(M)$ the algebra of $r$-forms on the manifold and $\mathfrak{g}$ the Lie algebra of the structure group, we have, locally,
\begin{equation} \label{AandFdefinitions}
    \Omega^1(M)\otimes\mathfrak{g}\ni A = A_\mu^I \, \dd x^\mu \, T_I \,, \qquad \Omega^2(M)\otimes\mathfrak{g}\ni F=\tfrac{1}{2} F_{\mu\nu}^I \, \dd x^\mu \wedge \dd x^\nu \, T_I \,,
\end{equation}
with $\{\dd x^\mu\}$  the coordinate  basis of one-forms, and $T_I$ a basis of generators of $\mathfrak{g}$. In a non-abelian gauge theory the Lie algebra generators do not commute, namely $[T_I,T_J]=c_{IJ}^K T_K$ with $c^K_{IJ}$  the \textit{structure constants}. When all the structure constants are zero, as is the case for electrodynamics, the theory is called abelian. From Eq. \eqref{AandFdefinitions}, it is clear that the gauge potential $A$ is a one-form and is associated with the {\it connection one-form}, while the field strength $F$ is the {\it curvature  two-form}
 \cite{nakahara2018geometry}. The  curvature form is obtained from  the connection form through the \textit{covariant exterior derivative}, which is defined on Lie-algebra-valued forms as follows:
\begin{equation}
    \DD: \eta\in \Omega^r(M)\otimes \mathfrak{g}\to \DD \eta=\dd \eta+\tfrac{1}{2}[A, \eta] \in \Omega^{r+1}(M)\otimes \mathfrak{g}\,,
\end{equation}
with $[\zeta,\eta]= \zeta\wedge\eta- (-1)^{pq}\eta\wedge\zeta$, for $\zeta, \eta $ Lie-algebra-valued p- and  q-forms, respectively \cite{nakahara2018geometry}.
Thus,  $F=\DD A= \dd A+ A\wedge A$.  
On explicitly computing the curvature two-form $F$ of 
 Eq.~\eqref{AandFdefinitions} we find that
\begin{equation}
    \begin{aligned}
        F & = \dd A +A\wedge A \\
        & = \partial_\mu A_\nu^I \, \dd x^\mu \wedge \dd x^\nu \, T_I + A_\mu^J A_\nu^K  \, \dd x^\mu \wedge \dd x^\nu \, T_J T_K \\
        & = \tfrac{1}{2} \left( \partial_{[\mu} A_{\nu]}^I + A_{\mu}^J A_{\nu}^K \, c_{JK}^I \right) dx^\mu \wedge dx^\nu \, T_I.
    \end{aligned}
\end{equation}
Therefore, we recover the well-known field-strength components of non-abelian gauge theories in the following form:
\begin{equation}
    F^I_{\mu\nu} = \partial_\mu A^I_\nu - \partial_\nu A^I_\mu + c_{JK}^I \, A_{\mu}^J A_{\nu}^K \,,
\end{equation}
and, in the abelian case, the electromagnetic field  $F_{\mu\nu}=\partial_\mu A_\nu - \partial_\nu A_\mu$ \cite{nakahara2018geometry}. Let us notice that analogously to the transformation of affine connections under linear transformations, the connection one-form is not covariant under gauge transformations, but transforms in a non-homogeneous manner,
\begin{equation}\label{gaugeA}
    A'= g A g^{-1} -\dd g g^{-1}\,.
    \end{equation}
Then, it can be checked that, precisely because of this, the curvature two-form is gauge covariant
\begin{equation}
    F'= g F g^{-1}\,.
\end{equation}
Finally,  the gauge connection allows for the introduction of a covariant derivative on charged matter fields, exactly in the same way as the affine connection defines a covariant derivative of tensor fields in Riemannian geometry, namely, it implements the parallel transport of matter fields, which behave tensorially under the internal structure group \cite{nakahara2018geometry}. Its local expression reads
\begin{equation}\label{nablader}
        \nabla_\mu \psi  = \del_\mu\psi + A^J_\mu \, \rho(T_J)   \psi \,,
\end{equation}
with $\rho(T_J)$ being the appropriate representation of the  Lie algebra $\mathfrak{g}$ on the matter field $\psi$. 

\subsection{Gauge  theories in the fiber bundle formalism}

The appropriate geometrical setting for gauge theories is the framework of fiber bundles. There, the distinguished nature of radiation fields which are responsible for the propagation of fundamental forces, and matter fields which are associated with the constituents of matter, is made very clear. Interactions are described in terms of gauge potentials, which are one-form connections of principal fiber bundles, together with their curvature two-forms, whereas matter fields are {\it sections} of the associated vector bundles. The interaction between matter and radiation is geometrically understood as a modification of the spacetime manifold so that ordinary derivatives are not appropriate anymore. A parallel transport is needed to compare matter fields at different points of spacetime, constructed in terms of the gauge connection, which yields a covariant derivative. 

Roughly speaking, a principal fiber bundle with structure group $G$ and base manifold  $M$ (the spacetime),  is a manifold $P$, locally built from the Cartesian product $U_i\times G$, with $U_i$ some open set of $M$, $\bigcup_i U_i= M$, also called a {\it local chart}. It has a projection map $\pi: P\rightarrow M$ such that $\pi^{-1} (x)= G$, also called the fiber of  $P$ at $x\in M$.  Differently from the projection map, a {\it section} of the principal bundle is a smooth map, generally only locally defined $\sigma_i: U_i\rightarrow \pi^{-1}(U_i)$, such that $\pi\circ\sigma_i = {\rm id}_{U_i}$. The bundle is trivial if it is globally endowed with the Cartesian product $P=M\times G$. In such a case the sections are globally defined. There is a global right action of the group $G$ on the bundle and a local action from the left. 

The gauge connection is a Lie-algebra-valued one-form on the principal bundle $P$,  $\omega\in \Omega^1(P)\otimes \mathfrak{g}$ with curvature $\mathcal{F}= \DD\omega \in \Omega^2(P)\otimes \mathfrak{g}$. The gauge potential $A$ and its curvature introduced in the previous section, are nothing but their local expressions in some open set of $M$, $U_i$. More precisely, $\sigma_i^* \omega= A_i, \; \sigma_i^* \mathcal{F}= F_i$, with  $\sigma_i^* :\Omega^r (P)\rightarrow \Omega^r (U_i)$, the so-called pull-back map through the section $\sigma_i$. It is then clear that the gauge potential $A_i$ and its curvature $F_i$ are globally defined on spacetime only if the bundle is trivial (namely, the section $\sigma_i$ is globally defined). For more details see for example \cite{nakahara2018geometry}. 

Gauge transformations of the gauge potential $A$, Eq.~\eqn{gaugeA}, are therefore understood as transformations of $A_i$ from one local chart $U_i\subset M$ to $A_j$ in another chart $U_j\subset M$, with $U_i\cap U_j\neq 0$ and $g: x\in U_i\cap U_j\rightarrow g(x)\in G$. The same holds for the gauge transformation of $F$. 

In order to describe matter fields we need vector bundles. Given a principal bundle $(P,\pi, M, G)$ encoding gauge fields, it is possible to construct  an associated vector bundle \cite{nakahara2018geometry}, namely a manifold $E$, locally diffeomorphic to the Cartesian product $U_i\otimes B$ with $U_i$ a local chart on the spacetime $M$ and $B$ a vector space carrying a representation of the structure group $G$ (hence, the group has a left action on the bundle, which is local). $B$ is called the typical fiber of the vector bundle. $E$ possesses a projection map $\pi: E\rightarrow M$ whose inverse $\pi^{-1}(x)\simeq B_x$ is the fiber of $E$ at $x\in M$. The sections of $E$, $s_i: x\in U_i\rightarrow s_i(x)\in \pi^{-1}(x)$, are vector-valued maps, with the property $\pi\circ s_i={\rm id}_{ U_i}$. 
These are the mathematical objects that describe matter fields, which are charged under (\textit{i.e.}, they carry a representation of)  the group $G$.
They can be locally described in terms of a base coordinate $x$ and a fiber coordinate $\psi(x)$,
\begin{equation}
    s_i(x)\sim (x, \psi(x))\,.
\end{equation}
Then, it is possible to derive Eq.~\eqn{nablader} as the local expression of the covariant derivative of the section $ s_i$ along a basis vector
\begin{equation}
 \nabla_\mu  s_i(x)\sim (x, \nabla_\mu \psi (x) )\,,
\end{equation}
with $\nabla_\mu$ determined by the gauge connection \cite{nakahara2018geometry}. 

In this geometric picture, gauge transformations are vertical automorphisms of the principal bundle, namely smooth maps $ \phi\,:\, P\to P$, which satisfy the $G$-equivariant condition,  $ \phi (pg)=\phi(p) g$ for all $p\in P, g\in G$ \cite{marathe2010topics}. Moreover, they are vertical, namely  $\pi \circ  \phi(p) =\pi(p)$, where $\pi$ is the projection map. Every $\phi\in Aut(P) $ induces a diffeomorphism $\tilde \phi$ on the basis manifold. The map $H$, which associates $\tilde\phi\in Di\!f\!f(M)$ to $ \phi\in Aut(P)$, is a group homomorphism. Thus, the kernel of $H$, given by those automorphisms of $P$ which are mapped to the identity in $Di\!f\!f(M)$, is a group. This allows for a mathematical definition of gauge transformations:
\begin{definition}[Gauge transformation] The gauge group of P is $\mathcal{G}(P):=ker(H)$. Its elements are called gauge transformations or vertical automorphisms.
\end{definition}
Bundle automorphisms of the principal bundle can act on every associated bundle, including matter/vector bundles, thereby defining automorphisms of the associated bundle \cite{hamilton2017mathematical}.

\section{Noncommutative gauge  and field theory} \label{sec:ncgft}

The study of noncommutative geometry emerged in the 1980s and remains an active area of research till today \cite{PMIHES_1985__62__41_0,Connes:2000by,Connes:2006ms,10.1007/3-540-53763-5_42,Dubois-Violette2001,gracia2000elements,landi1997introduction}. It has provided valuable insights into various areas of mathematics and physics, including gauge and field theories. Several approaches have been developed to deal with physics in a noncommutative spacetime. In these notes, we shall identify the latter with noncommutative algebras of functions defined on a commutative spacetime, with noncommutativity encoded in a suitable $\star$-product of the algebra, and we will mainly be concerned with gauge and field theory with such underlying noncommutative spacetime. A classical result of algebraic geometry states that the commutative algebra of continuous functions on a compact topological space provides an entirely equivalent description of the underlying space and its properties \cite{Gelfand}. Elaborating on this theorem, a noncommutative space is therefore {\it defined} as the dual object of a noncommutative algebra, with suitable properties. 
In other words, a noncommutative space is described through a dual noncommutative algebra,  the dual description being the only possible one, as the concept of a point is no longer available, unlike in its commutative counterpart \cite{Masson:2012yr}.

\subsection{Algebras}

Let us begin by recalling the formal definition of an algebra \cite{nakahara2018geometry}.
\begin{definition}[Algebra] \label{def:algebra}
    An \textit{algebra} $(\mathcal{A},+,*,\cdot)$ over a field $K$ is a set $\mathcal{A}$ equipped with three operations:
    \begin{itemize}
        \item[i.] An internal sum, $+\,:\, (a,b)\in \mathcal{A}\times\mathcal{A}\to a+b\in \mathcal{A}$, such that $(\mathcal{A},+)$ forms an abelian group. This means that the sum is associative $(a+b)+c=a+(b+c)$, commutative $a+b=b+a$, there exists an identity element with respect to the sum $a+e=a=e+a$, and each element has an inverse with respect to the sum $a+a^{-1}=e=a^{-1}+a$.
        \item[ii.] An external product, $*\,:\, (\zeta,a)\in K\times\mathcal{A}\to \zeta * a \in \mathcal{A}$, such that there exists an identity element with respect to it, and it is compatible with the internal product of $K$, namely $(\zeta\mu)*a=\zeta*(\mu*a)$. Moreover, the external product is distributive with respect to the internal sum of $\mathcal{A}$, namely $\zeta*(a+b)=\zeta*a+\zeta*b$, and the same holds for the internal sum of $K$.
    \end{itemize}
    The structure $(\mathcal{A},+,*)$ defined up to this point is a \textit{vector space} over the field $K$. It becomes an algebra by incorporating the third operation:
    \begin{itemize}
        \item[iii.] An internal product, $\cdot\,:\, (a,b)\in \mathcal{A}\times\mathcal{A}\to a\cdot b\in \mathcal{A}$, such that it is distributive with respect to the internal sum of $\mathcal{A}$ and is compatible with the external product. 
    \end{itemize}
    When the internal product is also associative, we refer to it as an associative algebra. Similarly, when the external product exhibits commutativity, we have a  commutative algebra. In the case where the product satisfies both properties, we have what is called a commutative associative algebra. 
\end{definition}
In the dual description of spacetime $M$, the latter is replaced by the associative, commutative algebra of smooth functions,  $(\mathcal{F}(M),\cdot)$ where $\cdot$ denotes the standard pointwise product $f\cdot g (x)= f(x) g(x)$. Therefore, a ``noncommutative spacetime'' replacing the smooth manifold $M$ will be an associative noncommutative algebra $(\mathcal{F}(M),\star)$ with  $\star$ being an associative and noncommutative product. Different  $\star$-products will yield different models of noncommutativity. A desirable property for physical applications is that the $\star$-product depends on some noncommutativity parameter, which can be set to zero to recover commutative spacetime, in the same way as $\hbar\rightarrow 0$ yields back classical mechanics from quantum mechanics.

\subsection{Modules as  generalization of vector spaces}

In the context of noncommutative spaces where the concept of points is no longer applicable, the notion of vector fields also becomes meaningless. Vector fields are smooth maps $X\,:\,x\in M\to X|_x\in T_xM$, which associate to each point in the spacetime a tangent vector at that point. However, in the absence of points, we need to consider a more general framework. 
\begin{definition}[Ring]
    A \textit{ring} $(R,+,\cdot)$ is a set $R$ equipped with two operations:
    \begin{itemize}
        \item[i.] An internal sum, $+\,:\, (a,b)\in R\times R\to a+b\in R$, such that $(R,+)$ forms an abelian group.
        \item[ii.] An internal product, $\cdot\,:\, (a,b)\in R\times R\to a\cdot b\in R$, such that $(R,\cdot)$ forms a semigroup, namely the product is associative. Moreover, the product is distributive with respect to the sum. 
    \end{itemize}
\end{definition}
If the internal product is commutative, then $(R,+,\cdot)$ is a commutative ring. It is worth noting that if $(R,\cdot)$ has additional properties that allow it to form an abelian group, then $(R,+,\cdot)$ becomes a field. In other words, a field is a more comprehensive structure than a ring, making the latter less restrictive.

\begin{definition}[Module]
    A \textit{left $R$-module} $(\mathbb{M},+,*)$ over a ring $R$, is a set $\mathbb{M}$ equipped with two operations:
    \begin{itemize}
        \item [i.] An internal sum $+\,:\,(a,b)\in \mathbb{M}\times \mathbb{M} \to a+b \in \mathbb{M}$, such that $(\mathbb{M},+)$ forms an abelian group.
        \item[ii.] An external product, $*\,:\, (\zeta,a)\in R\times \mathbb{M}\to \zeta * a \in \mathbb{M}$, such that there exists an identity element with respect to it, and it is compatible with the internal product of $R$. Moreover, the external product is distributive with respect to the internal sum of $\mathbb{M}$, and the same holds for the internal sum of $R$.
    \end{itemize}
    We can define a \textit{right $R$-module} in a similar manner, but with the distinction that all the external products are computed by multiplying the elements of the ring from the right instead of the left. If $R$ is a commutative ring, there is no distinction between a left and a right $R$-module, and it is simply referred to as an $R$-module.
\end{definition}
These operations are the same as those introduced in Definition \ref{def:algebra}, the only difference being that  $R$ is a ring and not a field, therefore modules are a generalization of vector spaces \cite{anderson2012rings}. The set of smooth functions over a manifold $\mathcal{F}(M)$, equipped with the pointwise product is a ring. It is also a module with respect to left/right multiplication by real or complex numbers. 
The set of vector fields $\mathfrak{X}(M)$, equipped with an internal sum $+\,:\,\mathfrak{X}(M) \times \mathfrak{X}(M) \to \mathfrak{X}(M)$ and an external product $*\,:\,\mathcal{F}(M) \times \mathfrak{X}(M) \to \mathfrak{X}(M)$ over the ring of smooth functions, $(\mathfrak{X}(M),+,\cdot)$ with
\begin{equation}\label{op}
    (X+Y)(p)\coloneqq X(p)+Y(p) \,, \qquad (X*f)(p)\coloneqq X(p)f(p) \,.
\end{equation}
is easily checked to be a left $\mathcal{F}(M)$-module, but not a right $\mathcal{F}(M)$-module (as the vector fields would act on functions on their right).

Matter fields, namely sections of vector bundles that carry a representation of the structure group associated with the gauge symmetry,  can be equivalently described as right modules over the ring of smooth functions on spacetime \cite{nakahara2018geometry}, a picture which can be easily generalized to the noncommutative setting \cite{Dubois-Violette1988}.

\subsection{Gauge transformations, connection, and curvature}

In Section \ref{sec:cgft} we discussed gauge transformations and their usual definition as maps from the spacetime manifold to the structure group. Within the framework of fiber bundles, gauge transformations are understood as automorphisms of a principal bundle which induce automorphisms of the associated bundles. When replacing vector bundles with right  modules, gauge transformations are seen as automorphisms of the module, and this definition naturally generalizes to the noncommutative case, as we shall see in Section \ref{DSec2}.  In Section \ref{sec:cgft}, we also provided a formal definition of connection and curvature within the fiber bundle formalism. A gauge connection on a principal bundle induces a covariant derivative on sections of the associated vector bundles, which correspond to matter fields. On replacing vector bundles with right modules over an algebra, the covariant derivative shall be defined in terms of a Koszul connection \cite{Dubois-Violette:1988cpg}, which in turn requires an appropriate differential calculus.  This will be the subject of the next section.
Then, in Section \ref{DSec1}, we will review the definition of connection and curvature in a noncommutative space. We will focus on the Moyal case, although the same procedure can be applied to other noncommutative spaces.

\section{Differential calculus} \label{sec:differential calculus}

In order to describe the dynamics of fields, it is necessary to introduce a differential calculus that has to be compatible with spacetime noncommutativity. To understand the relevance of the problem, let us consider the \textit{Kontsevich star-product} \cite{Kontsevich:1997vb}
\begin{equation} \label{Kont}
    f \star g \coloneqq f\cdot g+\tfrac{i}{2}\Theta^{\mu\nu}(x)\,\partial_\mu f\, \cdot \partial_\nu g + O(\Theta^2)  \,,
\end{equation}
with $\cdot$ being the ordinary pointwise product of functions (which will be omitted from now on) and $\Theta$ an antisymmetric matrix encoding the noncommutativity.  It is immediate to check that if $\Theta$ depends on spacetime coordinates,  ordinary derivatives are not derivations of the $\star$-product, namely they are not \textit{star-derivations}, as they violate the Leibniz rule,
\begin{equation} \label{Lei}
    \begin{aligned}
        \partial_\sigma(f \star g) &= (\partial_\sigma f)g+f(\partial_\sigma g) + \tfrac{i}{2}\partial_\sigma\Theta^{\mu\nu}(x)\,\partial_\mu f\, \partial_\nu g + \text{(higher orders)}\\
        &=(\partial_\sigma f)\star g+f\star (\partial_\sigma g) + \tfrac{i}{2}\partial_\sigma\Theta^{\mu\nu}(x)\,\partial_\mu f\, \partial_\nu g + \text{(higher orders)}\\
        &\neq (\partial_\sigma f)\star g+f\star (\partial_\sigma g) \,.
    \end{aligned}
\end{equation}
In what follows we consider two main examples:
\begin{example}[Canonical noncommutativity] \label{Ex:Moyal}
    The \textit{Moyal star-product} may be regarded as a specific instance of the Kontsevich star-product, which arises when $\Theta^{\mu\nu}$ is a constant matrix. It was introduced by \cite{Moyal:1949sk, Gronewold} in the context of Weyl quantization. It results  in the  noncommutative algebra $(\mathcal{F}(\mathbb{R}^n), \star_\theta)$, which we shall indicate with  with $\mathbb{R}^n_\theta$.  This results in a constant noncommutativity referred to  as \textit{canonical} or \textit{quantum-mechanics-like} noncommutativity (see Eq.~\eqref{qpnc}) described by the star-product:
    \begin{equation}\label{moyprod}
        f\star g= f\exp\left(\tfrac{i}{2}\Theta^{\mu\nu}\overleftarrow{\del_\mu}\,\,\overrightarrow{\del_\nu}\right)g,
        \end{equation}
        yielding for coordinate functions
    \begin{equation} \label{ncmoyal}
        [x^\mu,x^\nu]_\star=x^\mu\star x^\nu-x^\nu\star x^\mu =i\,\Theta^{\mu\nu}\,.
    \end{equation}
    We will review the Moyal algebra and its product in Section \ref{ScalarF_on_moyal} in more detail.  
    It is easily checked that, in this case, ordinary derivatives not only are derivations of the pointwise product, but they are also star-derivations. Indeed, when using Eq.~\eqn{moyprod} to compute the star commutator  $x^\nu\star f- f\star x^\nu$ we can verify that
    \begin{equation}\label{Moyalstarder}
        \del_\mu f= (\Theta^{-1})_{\mu\nu}[x^\nu,f]_\star\,.
    \end{equation}
    Therefore, the Leibniz rule is automatically satisfied because of the associativity of the star-product: 
    \begin{equation}
        [x, f\star g]_\star = [x, f]_\star \star g+ f\star[x,g]_\star\,.
    \end{equation}
    Derivations that are realized as commutators are called {\it inner} derivations. Therefore, inner derivations defined in terms of star-commutators,  are automatically star-derivations. 
\end{example}

\begin{example}[Lie-algebra type noncommutativity]
    A  less trivial kind of noncommutativity is \textit{Lie-algebra noncommutativity}, where the noncommutative matrix is linear in the coordinate functions. In this case, the star-commutator of the coordinate functions is given by
    \begin{equation}\label{lietype}
        [x^\mu,x^\nu]_\star = \tensor{c}{^\mu^\nu_\sigma}\,x^\sigma\,,
    \end{equation}
    where $\tensor{c}{^\mu^\nu_\sigma}$ are the structure constants of a Lie algebra. Ordinary derivatives $\del_\mu$ cease to be star-derivations, but similarly to the previous case, we do have a family of star-derivations, given by  inner derivatives:
    \begin{equation}\label{lietypeder}
        D_\mu f \coloneqq k[x^\mu,f]_\star\,,
    \end{equation}
    where $k$ is a suitable dimensional constant. In the commutative limit, they do not reproduce ordinary derivatives, as we shall see in Section \ref{DSec1} where an example of linear noncommutativity is considered in detail. 
    \end{example}
If we are to interpret spacetime noncommutativity as a relic of a yet-to-be-achieved quantum gravity theory, a desirable characteristic of noncommutative field theories and noncommutative dynamical systems in general,  is that they certainly should reproduce standard theories in the commutative limit $\Theta\rightarrow 0$ (see for example \cite{Kupriyanov:2020sgx} for a discussion on this issue).  Definitely,  the issue of a well-defined differential calculus that is not only mathematically sound but also physically adequate, is an important one. Different proposals in this regard have been made. 

One important approach involves employing a twisted differential calculus for noncommutative spaces whose star-product is defined in terms of a twist operator \cite{Vassilevich:2006tc,Aschieri:2006ye, Chaichian:2006wt, Aschieri:2007sq}, as we will see in Section \ref{DSec3}. Another approach is the so-called derivation-based differential calculus, which we are going to review in the following.

\subsection{Derivation-based differential calculus} \label{derba}

The concept of derivation-based differential calculus was introduced in the context of noncommutative geometry by Dubois--Violette and Michor in their publications \cite{Dubois-Violette1988, Dubois-Violette:1994dsm,Dubois-Violette1996}. For a more general perspective see  \cite{Segal67, Segal70, Landi:1989qh}. This is an algebraic definition of differential calculus that can be applied to any associative algebra, whether commutative or not. For mathematical details and applications to noncommutative field theory also see \cite{Marmo:2004re, Cagnache:2008tz, deGoursac:2008bd, Wallet:2008bq, MASSON_2008, Marmo:2018gls} The following review is based on \cite{Marmo:2018gls}.

Given an orientable  $n$-dimensional differentiable manifold $M$, it is well known that the differential calculus on it is the differential graded algebra $(\dd, \Omega(M)=\oplus_{k=0}^N\Omega^k(M))$, with $\Omega^{k}(M)$ the set of $k$-exterior forms and $\dd:\Omega^k(M)\to\Omega^{k+1}(M)$ the (graded)  exterior derivative\footnote{The exterior derivative is formally defined as an antiderivation. In general, a linear map $\DD$ is referred to as a \textit{graded derivation of degree $k$} if it maps $\Omega^r\to\Omega^{r+k}$ and satisfies the graded Leibniz rule: $\DD(ab)=\DD(a)\,b+(-1)^{k|a|}a\,\DD(b)$, where $|a|$ denotes the rank of $a$. When $k$ is odd, $\DD$ is specifically called an \textit{antiderivation}. Hence, according to this definition, the exterior derivative is an antiderivation of degree $k=+1$  with $\dd^2=0$ \cite{Michor1989}.\label{note:antiderivation}}. The commutative algebra of smooth functions on $M$ is identified with the zero forms, $\mathcal F(M)=\Omega^0(M)$. The $\mathcal F(M)$-bimodule\footnote{Namely, left and right $\mathcal F(M)$-module.} of one-forms is dual to the set of vector fields $\mathfrak{X}(M)$. The set $\mathfrak X(M)$ coincides  with the space of all derivations of $\mathcal F(M)$; it is an infinite-dimensional Lie algebra, with the Lie bracket provided by the commutator   $[X,Y]f=X(Yf)-Y(X f)$ (with $X, Y\in\mathfrak X(M)$ and $f\in\mathcal F(M)$). As we have already mentioned, $\mathfrak X(M)$ is also a left $\mathcal F(M)$-module, namely, given a vector field $X$, the product $f \, X$ is again a vector field, hence a derivation of the algebra of functions; this is an important difference with respect to the noncommutative case, as we will see below. 

The differential calculus can be defined algebraically by means of the duality between one-forms and vector fields, now seen as derivations. Since the construction is similar, we shall illustrate the construction in detail for the noncommutative case, the commutative case being trivially obtained on replacing star derivations with derivations, and $\mathcal{A}$-modules with $\mathcal F(M)$-modules. We shall highlight the differences when relevant. When the algebra $\mathcal F(M)$ is replaced by a  noncommutative algebra $\mathcal A$, the problem of defining a differential calculus is to be addressed within the  Gelfand duality \cite{Gelfand}. It has been widely studied, yielding different proposals,  as we already mentioned. Within the approach considered here, one  starts from  a (finite-dimensional) Lie algebra of derivations acting on $\mathcal A$. The latter  dually defines a $\mathcal A$-bimodule of forms\footnote{Note that in the noncommutative case, left and right modules do not in general coincide.} and a whole  differential graded algebra that we interpret as a differential calculus on $\mathcal A$. 

Let us assume indeed (see \cite{Segal67, Segal70, Landi:1989qh, Dubois-Violette2001})  that    $\mathfrak g$ is  a Lie algebra acting upon an associative algebra with unity, $(\mathcal A,\star)$, by derivations, \textit{i.e.}, $\rho:\mathfrak g\to\,\mathrm{End}(\mathcal A)$ is a linear map with $[\rho(X_a),\rho(X_b)]=\rho([X_a,X_b])$ and 
 $\rho(X)(f\star g)\,=\,(\rho(X)f)\star g+f\star (\rho(X)g)$ for any $X,X_a,X_b\in\mathfrak g$ and $f,g$ in $\mathcal A$. Let us denote by 
$C_{\wedge}^n(\mathfrak g, \mathcal A)$  the set\footnote{Differently from the commutative case,  the space of derivations for a given noncommutative algebra $\mathcal A$ is a left module only with respect to the center $Z(\mathcal A)$ of the algebra, since $(f\star X) (g\star h)\ne f\star \rho(X) (g)\star h + g \star f\star \rho(X)(h) $, unless $f \in Z(\mathcal A)$.} of $Z(\mathcal A)$-multilinear alternating maps $\omega\,:\,X_1\wedge\dots\wedge X_n\,\mapsto\,\omega(X_1, \dots, X_n)$ from $\mathfrak g^{\otimes n}$ to $\mathcal A$. Let us consider then the  graded vector space $C_{\wedge}(\mathfrak g, \mathcal A)=\oplus_{j=0}^{j=\mathrm {dim}\,\mathfrak g}C_{\wedge}^n(\mathfrak g, \mathcal A)$, with $C_{\wedge}^0(\mathfrak g, \mathcal A)=\mathcal A$. We can define a wedge product by
\begin{equation}
\label{wedp}
(\omega\wedge\omega')(X_1,\dots,X_{k+s})=\frac{1}{k!s!}\sum_{\sigma\in \mathcal S_{k+s}}\mathrm{sign}(\sigma)\,\omega(X_{\sigma(1)},\dots,X_{\sigma(k)})\, \omega'(X_{\sigma(k+1)},\dots,X_{\sigma(k+s)})\,,
\end{equation}
where $\omega\in C_{\wedge}^k(\mathfrak g, \mathcal A),\,\omega'\in C_{\wedge}^s(\mathfrak g, \mathcal A)$, $X_j\,\in\,\mathfrak g$, and  $\mathcal S_{k+s}$ is the set of permutations of $k+s$ elements.
Thus we define the operator $\dd:C_{\wedge}^n(\mathfrak g, \mathcal A)\to C_{\wedge}^{n+1}(\mathfrak g, \mathcal A)$  by 
\begin{align}
    \dd\omega\, (X_0, X_1,\dots, X_n)&=\sum_{k=0}^n(-1)^k\rho(X_k)\omega(X_0,\dots,\hat{X}_k,\dots,X_n)\\
    & \qquad+\frac{1}{2}\sum_{r,s}(-1)^{k+s}\,\omega([X_r,X_s], X_0,\dots, \hat{X}_r,\dots,\hat{X}_s,\dots, X_n)\,, \label{ddef}
\end{align} 
with $\hat{X}_r$ denoting that the $r$-th term is omitted. Such operator  is easily proven to be a graded antiderivation with $\dd^2=0$, so $(C_{\wedge}(\mathfrak g, \mathcal A), \dd)$ is a graded differential algebra. Operatively, it is constructed by starting with zero-forms (the elements of $\mathcal A$) and the operator $\dd$ to build $C_{\wedge}^1(\mathfrak g, \mathcal A)$ as a left (or right) star-module,
\begin{equation}
C_{\wedge}^1(\mathfrak g, \mathcal A)\ni \omega= \sum f_j\, \dd g_j (X) = f_j\star \rho(X) g_j\,,
\end{equation}
with higher forms obtained iteratively, \textit{e.g.,} for two-forms  $ \alpha = \sum f_i\, \dd \omega_i$, $\omega_i\in C_{\wedge}^1(\mathfrak g, \mathcal A)$.

Although the relations \eqref{wedp} and \eqref{ddef} are valid for both commutative and noncommutative algebras, with $Z(\mathcal A)$ replaced by the whole $\mathcal F(M)$ in the commutative case, when the algebra $\mathcal A$ is not commutative one easily sees that it is  in general $f_1\dd f_2\neq (\dd f_2)f_1$ and $\omega\wedge\omega'\neq (-1)^{k s}\omega'\wedge\omega$. One has indeed 
$(f_1\dd f_2)(X)=f_1\star(\rho(X)f_2)$ and $((\dd f_2)f_1)(X)=(\rho(X)f_2)\star f_1$, while, for $\omega, \omega'$ one-forms,  $(\omega\wedge\omega')(X_1,X_2)=\omega(X_1)\star \omega'(X_2)-\omega(X_2)\star \omega'(X_1)$ and $(\omega'\wedge\omega)(X_1,X_2)=\omega'(X_1)\star \omega(X_2)-\omega'(X_2)\star \omega(X_1)$.
This exterior algebra is an example of a derivation-based calculus, where the derivations come from the action of the Lie algebra $\mathfrak g$ upon the algebra $\mathcal A$. 
By construction, every element in $C_{\wedge}(\mathfrak g,\mathcal A)$
can be written as a sum of $a_0\, \dd a_1\wedge\dots\wedge\dd a_n$ terms with $a_j\in\mathcal A$.

Upon the graded differential algebra $C_{\wedge}(\mathfrak g,\mathcal A)$ a contraction operator $\iota_X$ can be defined. If $X\in\mathfrak g$, then 
\begin{equation}\label{contX}
\iota_{X}\omega\, (X_1,\dots,X_n)\,=\,\omega(X,X_1,\dots,X_n)\,, \qquad X_j\in\mathfrak g,
\end{equation}
gives a degree $(-1)$ antiderivation from $C_{\wedge}^{n+1}(\mathfrak g, \mathcal A)\to C_{\wedge}^n(\mathfrak g,\mathcal A)$. The operator defined by $\mathcal{L}_X=\iota_X\dd+\dd \, \iota_X$ is the degree zero Lie derivative along $X$, and the set $(C_{\wedge}(\mathfrak g, \mathcal A), \dd, \iota_X, \mathcal{L}_X=\iota_X\dd+\dd \iota_X)$ gives a Cartan calculus on $\mathcal A$ depending on  the Lie algebra $\mathfrak g$ of derivations.

When one is concretely searching for derivations of a given noncommutative algebra, in view of physical applications, two other important properties are required:
\begin{itemize}
    \item[i.] Derivations must be independent. A set of derivations is considered independent if any linear combination of them, with functions as coefficients, is nonzero unless all the coefficients are zero everywhere. As an example, let us consider the commutative algebra of smooth functions  on  $\mathbb{R}^3$, and  the set of vector fields  $Y_i=\tensor{\varepsilon}{_i_j^k}x^j \partial_k$. They belong to the left module  of  derivations of $\mathcal F(\mathbb{R}^3)$, namely  $f\, Y_i$ is also a derivation, but they are not independent, since it is  easily checked that the combination   $x^iY_i=0$ for $x^i\ne 0$. This means in particular that they cannot generate the whole module of derivations for $\mathcal F(\mathbb{R}^3)$, namely, they are not enough to retrieve the standard differential calculus of $\mathbb{R}^3$.
    \item[ii.] Derivations must be sufficient, meaning that only  constant functions are annihilated by all of them. In the previous example, functions which depend on $r=\smash[b]{\bigl(\sum_i {(x^i)}^2}\bigr)^{1/2}$ are annihilated by all $Y_i=\tensor{\varepsilon}{_i_j^k}x^j \partial_k$, but they are certainly not constant. Therefore,  this set of derivations is neither independent nor sufficient.
\end{itemize}

These two conditions trivially extend to the noncommutative case. When derivations are inner, namely, they are realized through star-commutators, the request that they are sufficient is equivalent to requiring that only elements that belong to the center of the algebra are in the kernel of the module of derivations. The center of the algebra, in turn, should only contain constant functions. This request is important to recover the right commutative limit, as we shall see in the coming sections when considering Lie-algebra type noncommutativity.

Returning to the example of derivations in $\mathbb{R}^3$, if we consider a Lie-algebra type noncommutativity, as in Eq.~\eqn{lietype}, with $\tensor{c}{^{ij}_k}=\tensor{\varepsilon}{^{ij}_k}$, the star-commutators $[x^i, \cdot]_\star:= D_i$ are star-derivations. They are independent because the only linear combination with nonzero functions that adds to zero is given by $x^i \star D_i (f) + D_i (f) \star x^i=0$. But the product $x^i D_i$ is not a derivation because $x^i$ is not in the center of the algebra. However, these derivations are not sufficient, because functions of $\sum_i (x^i)^2$ are in the center of the algebra, therefore they are annihilated by $D_i, i= 1, \dots 3$ but they are not constant functions. We shall see in Section \ref{DSec1} what are the physical consequences of this observation.

\section{Noncommutative scalar field theory on Moyal space}
\label{ScalarF_on_moyal}

In Section \ref{sec:ncgft} we discussed various examples of star-products, including the Moyal star-product for the noncommutative algebra of functions $\mathcal A= \mathbb{R}^{k}_\theta$, introduced in Example \ref{Ex:Moyal}. The latter   gives rise to canonical noncommutativity or quantum-mechanics-like noncommutativity, with  the star-commutator between the coordinate functions of the associated noncommutative space, known as the \textit{Moyal space},  given by
\begin{equation}\label{Moyalcomm}
[x^\mu,x^\nu]_\star=i\Theta^{\mu\nu} \, .
\end{equation}
This is exactly the canonical phase-space noncommutativity if $k$, the dimension of spacetime, is even, and the spacetime coordinates $(x^j, x^{j+\frac{k}{2}})$, with  $j\in (1,\dots \frac{k}{2})$,  are replaced by  $(q^j, p_j)$. Indeed, the Moyal space is modeled on the phase space of quantum mechanics. In the WWM formulation, as discussed in Section \ref{phasespace}, classical observables undergo a process of quantization (Weyl quantization), where phase-space coordinates $q$ and $p$ are transformed into noncommuting operators $\hat{q}$ and $\hat{p}$. In order to have a classical-like description (Wigner picture)  a process of dequantization takes place, where quantum observables are converted into operator symbols, which are smooth functions on the phase space, equipped with a noncommutative  product (the Moyal product). This noncommutative algebra of functions over the phase space provides an equivalent description of quantum mechanics. A similar approach can be applied to spacetime, resulting in a constant noncommutativity between spacetime coordinates. In the following, we give a more formal definition of the Moyal algebra. 

\subsection{The Moyal algebra}

Consider a finite dimensional space $\mathbb{R}^k$ (Euclidean for simplicity) and an antisymmetric $k\times k$ matrix, $\Theta$. Let $f,g\in \mathcal{S}(\mathbb{R}^k)$ be Schwartz functions, which are smooth and rapidly decreasing. The Moyal product is defined as follows (see the appendix of \cite{Gracia-Bondia:2001ynb} for a review):
\begin{equation}\label{Moyint}
    (f\star_{\Theta}g) (x)\coloneqq \tfrac{1}{(2\pi)^k}\int\! \dd^k u\, \dd^k v\, f(x-\tfrac{1}{2}\Theta u)\, g(x+v)e^{-i\,uv} \,.
\end{equation}
If the antisymmetric matrix $\Theta$ is nondegenerate, then the space must be even-dimensional, $k=2n$. By making the invertible change of variables $s=-\tfrac{1}{2}\Theta u$, and defining $\theta>0$ as $\theta^{2n}\coloneqq \operatorname{det} \Theta$, we can rewrite the Moyal product as
\begin{equation}\label{moyalint}
    (f\star_{\Theta}g) (x)=\tfrac{1}{(\pi\theta)^{2n}}\int \! d^{2n}s \, d^{2n}v \, f(x+s)\,g(x+v) e^{-2i\Theta^{-1}sv}\,,
\end{equation}
which is a familiar expression in the classical-like description of quantum mechanics \cite{Gayral:2003dm}.  $(\mathbb{R}^{2n}, \Theta^{-1}) $ is a symplectic vector space with  the inverse $\Theta^{-1}$ being  the canonical symplectic two-form  $\omega= \tfrac{1}{2} \Theta^{-1}_{\mu\nu} dx^\mu \wedge dx^\nu$, or, equivalently, $\Lambda= \tfrac{1}{2} \Theta^{\mu\nu} \del_\mu \wedge \del_\nu$ the non-degenerate Poisson bracket. In  Darboux coordinates,  $\Theta$ and its inverse are thus represented  by
\begin{equation}
    \Theta^{\mu\nu}=\left( \begin{array}{cc}
    0 & -\mathbb{I}_n \\
    \mathbb{I}_n & 0
    \end{array} \right) \,, \qquad \Theta_{\mu\nu}=\left( \begin{array}{cc}
    0 & \mathbb{I}_n \\
    -\mathbb{I}_n & 0
    \end{array} \right) \,,
\end{equation}
where $\mathbb{I}_n$ denotes the $n\times n$ identity matrix. It can be checked that the popular expression already introduced for the Moyal product, Eq.~\eqn{ncmoyal}, is nothing but the asymptotic expansion of its integral formula, Eq.~\eqn{Moyint}.  

The zeroth order of the Moyal product corresponds to the commutative product, while the first and higher orders  involve the first and higher derivatives of the functions under consideration. It is therefore highly non-local.   Once more, when computing the Moyal star-commutator of coordinate functions, one retrieves canonical, or quantum mechanics-like, noncommutativity. 

As discussed in Section \ref{DFR argument}, a non-trivial commutator between coordinate functions leads to the existence of a minimal area in spacetime. The Moyal space, being the simplest example of noncommutativity, has been extensively studied in various contexts, including field theories, gauge theories, condensed matter theories, and more.

\subsection{Differential calculus on the Moyal algebra}

In order to define a derivation-based differential calculus for the Moyal space, according to Section \ref{sec:differential calculus}, it is necessary to define derivations. We have already seen that standard derivatives $\del_\mu$, $\mu=1,\dots,k$, are at the same time derivations of the commutative algebra $\mathcal{F}(M)$ and star-derivations of the noncommutative algebra $\R^k_\theta$. They give a representation of the Lie algebra of translations in $k$ dimensions, $\mathfrak t$. For even values of $k$, there is a larger algebra with this property, it is the Lie algebra of the inhomogeneous symplectic group $ISp(k,\mathbb{R})$, which   consists of translations and real symplectic transformations of $\mathbb{R}^k$.  Such a large algebra of derivations has been employed in \cite{Marmo:2004re, Marmo:2018gls} to obtain a derivation-based differential calculus for Lie-algebra type noncommutative spaces, which can be realized as subalgebras of a suitable  Moyal algebra.  However, for the purpose of developing a differential calculus for the Moyal space, it is enough to consider the minimal algebra of derivations represented by translations, $\mathfrak{t}$. The generators of translations, denoted as $P_\mu$, are given by
\begin{equation}
    \rho(P_\mu)\coloneqq \partial_\mu = -{i} D_\mu = -i (\Theta^{-1})_{\mu\nu}[x^\nu,\cdot]_\star\,,
    \end{equation}
as in  Eq.~\eqref{Moyalstarder}, namely they are inner derivations. They are a left module over the center of the algebra, $Z(\R^n_\theta)$, therefore we have fewer derivations than in the commutative case, as expected. 

According to the definitions \eqn{ddef} and \eqn{contX}, the exterior derivative $\dd$  and the contraction operator $\iota_X$ are defined by duality, acting on translations. When acting on functions, $\dd$ operates as follows:
\begin{equation}
    \dd f(P_\mu)=\rho(P_\mu)(f)=-i (\Theta^{-1})_{\mu\nu}[x^\nu,f]_\star\,,
\end{equation}
while, when $\iota_X$ acts on the one-form $\alpha= g \dd f$, we have
\begin{equation}
    \iota_{P_\mu} \alpha= \alpha(P_\mu) = g \star \dd f(P_\mu)=g\star \rho(P_\mu) f= g\star \partial_\mu f \,.
\end{equation}
The generalization to higher forms is straightforward. For example,  for $\omega= f \dd g \wedge \dd h$ we have $(\iota_{P_\mu}\omega) (P_\nu)= \omega (P_\mu, P_\nu)= f\star (\del_\mu g\star \del_\nu h - \del_\mu h \star \del_\nu g)$.  

An additional significant property of the Moyal product is its cyclicity under integration, namely
\begin{equation}
    \int\!\dd\Omega\,  f_1\star \dots \star f_n= \int\!\dd\Omega\, f_i\star \dots \star f_n\star f_1\star\dots\star f_{i-1}\,, \qquad \forall \, n\,, \ i\leq n\,.
\end{equation}
This property is often referred to as the \textit{trace property}, as it characterizes the trace operation in linear algebra.
In particular, the Moyal product is {\it closed}, namely we have \cite{curtright2013concise}
\begin{equation}\label{closure}
    \int\!\dd\Omega\,  f_1\star f_2=\int\!\dd\Omega\, f_2\star f_1=\int\!\dd\Omega\, f_1\,f_2\,.
\end{equation}
The latter can be easily proven by using the asymptotic expansion of the product and repeatedly integrating by parts. Notice that other products yielding the same spacetime noncommutativity are cyclic but not closed (\textit{e.g.}, the Wick--Voros and $s$-ordered products \cite{Manko:2004syv, Lizzi:2014pwa}).
Cyclicity is an important requirement in noncommutative gauge theory, as it ensures gauge invariance of the action functional, and in general, it is a suitable requirement in noncommutative field theory because it guarantees that the action functional is a scalar, as we shall explore in more detail in the following section.

\subsection{Noncommutative \texorpdfstring{$g\varphi^{\star 4}$ scalar field theory on Moyal space}{}} \label{subsec:scalar_moyal}

In what follows, we will review an important and well-studied example of a field theory on $\R^n_\theta$. We consider the noncommutative analog of the Euclidean scalar field theory $g \varphi^4$ with a mass term, described by the action functional
\begin{equation}\label{nclp4}
    S_{NC}[\varphi]=\int\! \dd^4 x \, \left( \tfrac{1}{2}D^\mu \varphi \star D_\mu \varphi - \tfrac{1}{2} m^2 \varphi^{\star 2} -\tfrac{1}{4!}g\varphi^{\star 4} \right)\,,
\end{equation}
where the  notation $\varphi^{\star n}$ stands for $\varphi\star\dots\star \varphi$ $n$ times, and $\star$ is the Moyal product. 
The commutative case is retrieved when replacing the $\star$-product with the pointwise product. 
The first two terms in the action correspond to the kinetic and mass terms, respectively. When only these terms are present, we have the action of a free massive scalar field. Both terms are quadratic in the fields, therefore the closure property of the Moyal product  \eqn{closure} applies. Thus, the free action is the same for the commutative and noncommutative case. 
In the framework of quantum field theory, this implies that the Feynman propagator is identical in the two cases, namely
\begin{equation}
    \Delta_F(x-x')= \int\! \frac{\dd^4k}{(2\pi)^4} \,  \frac{ie^{-ik(x-x')}}{k^2+m^2}\,.
\end{equation}
Indeed, given its definition, in analogy with the commutative case,
\begin{equation} \label{Feydef}
    \begin{aligned}
       \Delta_F(x-x') &\coloneqq \langle 0| T \varphi(x) \star \varphi(x')|0\rangle \\
        &= \theta(t-t')\, \langle 0| \varphi(x) \star \varphi(x')|0\rangle + \theta(t'-t)\, \langle 0| \varphi(x') \star \varphi(x)|0\rangle\,,
    \end{aligned}
\end{equation}
with $T$ the time-ordering operator, 
\begin{equation}
    T \varphi(x) \star \varphi(x') =\begin{cases}
        \varphi(x)\star\varphi(x')\,, & t>t'\,,\\
       \varphi(x')\star\varphi(x)\,, & t<t'\,,
    \end{cases}
\end{equation}
and $\theta(t-t')$ the Heaviside step function, it is sufficient to observe that the  $\star$-product can be safely removed from \eqn{Feydef} because of the closure property \eqn{closure}, and the star-derivations $D_\mu$ are given by the ordinary derivatives. Then the computation of $\Delta_F(x-x')$ becomes the standard one, contained in any introductory textbook of quantum field theory (see for example \cite{lancaster2014quantum, peskin2018introduction}).  Therefore, the Feynman propagator remains the same in both the commutative and noncommutative cases, if the Moyal product is employed.

The third term of  the action  \eqref{nclp4}  is the self-interaction term, which involves three star-products. In the commutative case, the momentum-space Feynman rules provide a framework for calculating scattering amplitudes in field theory. According to these rules, each vertex contributes with a factor of $g$, and the momentum conservation must be imposed by including the delta function $(2\pi)^4\delta^4\bigl(\sum_{a=1}^4 k_a\bigr)$, where $k_a$ are the momenta of the incoming and outgoing fields. Finally, the Feynman rules prescribe that we integrate overall undetermined momenta using $\int\! \frac{\dd^4k}{(2\pi)^4}$ \cite{peskin2018introduction,lancaster2014quantum}. In the noncommutative case, the interaction term is modified, resulting in a deformed expression for the vertex, which presents an additional phase factor relative to the commutative theory (see Appendix \ref{appendix} for proof),
\begin{equation}
    V_\star=-i g \prod_{a<b} e^{-\tfrac{i}{2}k_a \wedge k_b} \,.
\end{equation}
Here we introduced the convenient notation $k_a\wedge k_b=k_{a \mu} \Theta^{\mu\nu} k_{b\nu}$, where Latin indices $a,b,\dots$ enumerate the momenta of different fields, and Greek indices $\mu,\nu\dots$ are spacetime indices. This star-vertex is therefore not invariant under arbitrary permutations of $k_a$, but it retains cyclic-permutation invariance. It is important to note that in the commutative limit, where $\Theta^{\mu\nu}\to 0$, the commutative vertex is recovered, \textit{i.e.}, $V_\star\to V$ \cite{Minwalla:1999px}. This is the only modification to the Feynman rules.

\begin{figure}[h]\center
\includegraphics[width=0.7\textwidth]{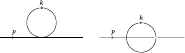} 
\captionsetup{width=.8\linewidth}
\caption{\footnotesize{Planar (on the left) and non-planar (on the right) one-loop corrections to the tree-level propagator in $g\varphi^4$ theory.}}\label{Fig:1loops}
\end{figure}

With the Feynman propagator and star-vertex at hand, we can now proceed to calculate loop corrections to the propagator (see Appendix \ref{appendix} for the explicit calculation). In the noncommutative case, the propagator receives one-loop corrections from two diagrams, one planar and the other non-planar (Fig.~\ref{Fig:1loops}),
\begin{equation}
    \Delta^{(1)}_{planar}=\frac{g^2}{3}\int\! \frac{\dd^4k}{(2\pi)^4}\frac{1}{k^2+m^2} \,, \qquad \Delta^{(1)}_{non-planar}=\frac{g^2}{6}\int\! \frac{\dd^4k}{(2\pi)^4}\frac{e^{i k \wedge p}}{k^2+m^2} \,,
\end{equation}

\noindent The planar diagram is proportional to the one-loop correction of the commutative theory, while the non-planar diagram differs from its commutative analog by a phase factor, which tends to 1 in the commutative limit \cite{Minwalla:1999px}. This new factor in the noncommutative case renders the one-loop diagrams finite in the UV regime. However, it also introduces new challenges in the IR regime, giving rise to the problem known as \textit{UV/IR mixing} \cite{Minwalla:1999px}. Various approaches have been proposed to address this issue, including the development of quantum field theories with alternative noncommutative products or the modification of the $g\varphi^4$ action through the addition of new terms \cite{Grosse:2003nw, Grosse:2004yu, Grosse:2005iz,  Gurau:2008vd}.

\section{Linear noncommutativity: the case of \texorpdfstring{$\mathbb{R}^3_\lambda$}{}} \label{DSec1}

Motivated by what we encountered in the case of quantum mechanics in Section \ref{phasespace}, it is clear that we have focused so far on the easiest example of noncommutativity in spacetime. This is the constant noncommutativity, which gives rise to the Moyal noncommutative space. In Section \ref{ScalarF_on_moyal}, we studied in detail an interacting scalar field theory formulated on the Moyal space. It is worth mentioning that even in the simplest examples of noncommutativity and field theory, the resulting scenario is not without its troubles. In this section, we will go beyond constant noncommutativity to deal with a different type of noncommutativity, known as Lie-algebra type or, more generally, linear noncommutativity. These cases are characterized by a star-commutator of coordinate functions, which replicates the Lie-bracket structure of classical Lie algebras. This \textit{linear noncommutativity} was thoroughly analyzed in \cite{Gracia-Bondia:2001ynb}, and various star-products were proposed to reproduce three-dimensional Lie algebras at the coordinate level. In this section, we will consider the $\mathfrak{su} (2)$ Lie algebra that gives rise to the \textit{noncommutative space $\mathbb{R}^3_\lambda$} introduced in \cite{Hammou:2001cc}.

\subsection{The algebra of \texorpdfstring{$\mathbb{R}^3_\lambda$}{} } 

\label{subsec:algebraR3}
In the following, we will show  how  to implement a $\mathfrak{su} (2)$-like noncommutativity on the Euclidean space $\mathbb{R}^3$ using a star-product in $\mathcal{F} (\mathbb{R}^3)$. This star-product ensures that the star-commutator of coordinate functions is
\begin{equation}
    [x_i, x_j]_\star = i \lambda \epsilon_{i j}^k x_k \ .
\end{equation}
In general, the construction of a non-formal star-product for a given algebra of functions starting from the commutator of the coordinate functions is not an easy task. In the case under consideration, the classical analog of the so-called \textit{Jordan--Schwinger map}\footnote{In physics, the Jordan--Schwinger map allows the realization of the $\mathfrak{su} (2)$ Lie algebra in terms of annihilation and creation operators of the two-dimensional harmonic oscillator \cite{Biedenharn1965}.} has been employed. The latter embeds the algebra of functions on $\R^3$ into the algebra of functions on $\R^4$ via a quadratic realization of the $\R^3$ coordinates. Therefore, the former is shown to be a $\star$-subalgebra of the latter, as we will shortly review below.

We will consider the so-called \textit{Wick-Voros star-product} \cite{Voros:1989iu}, which is a variation of the Moyal star-product that produces normal-ordered operators instead of symmetric-ordered ones. This star-product yields
\begin{equation}
     (\phi \star_V  \psi) (z_a, \Bar{z}_a) = \phi (z, \Bar{z}) \exp(\theta \overleftarrow{\partial}_{z_a} \overrightarrow{\partial}_{\Bar{z}_a}) \psi(z, \Bar{z}) \ , \quad a = 1, 2 \ ,
\end{equation}
where $z_a$ are complex coordinates in $\C^2\simeq\R^4$, with $z_1 = \frac{1}{\sqrt 2}(y_0 + i y_3)$ and $z_2 = \frac{1}{\sqrt 2}(y_1 + i y_2)$. 
The star-commutator for complex coordinate functions in this case yields
\begin{equation}
    [z_a, \Bar{z}_b]_\star = \theta \delta_{a b} \ ,
\end{equation}
with $\theta$ being a constant real parameter, namely the Moyal star-commutator of Eq.~\eqn{ncmoyal} for the real coordinates $y_\mu$\footnote{A comparison of Moyal and Wick--Voros star-products, as well as their corresponding noncommutative field theories, are investigated in \cite{Galluccio:2008wk}}. The Jordan--Schwinger map is then explicitly given by
\begin{equation}
    x_{\mu} = \frac{1}{2} \Bar{z}_a \sigma_{\mu}^{a b} z_b \ , \quad \mu = 0, \ldots, 3 \ , \label{JSm}
\end{equation}
with $\sigma_0$ the identity matrix, and $\sigma_i$ being the Pauli matrices. This map accomplishes the aforementioned embedding by identifying the coordinates in $\mathbb{R}^3$ as a quadratic combination of coordinates in $\mathbb{R}^4$ (represented as two copies of $\mathbb{C}^2$).

The above coordinate functions generate a subalgebra that is closed with respect to the Wick--Voros star-product, where one can check that $f\star_V g (x) \in \mathcal F(\R^3)$. 
Indeed, it is possible to obtain a closed expression for the induced star-product in $\R^3$, which reads
\begin{equation}
    (\phi \star_\lambda  \psi) (x) = \left. \exp \left[ \frac{\lambda}{2} \left( \delta_{i j}x_0 + i \epsilon_{i j}^k x_k \right) \frac{\partial}{\partial u_i} \frac{\partial}{\partial v_j} \right] \phi (u) \psi (v) \right|_{u = v = x} \ . \label{LNCsp}
\end{equation}
This implies that\footnote{We consider here $z_a$ to have length dimension $1/2$, so we choose $\lambda = \theta$, of length dimension $1$.}
\begin{equation}
    \begin{split}      
        x_i \star x_j &= x_i x_j + \frac{\lambda}{2} \left( \delta_{i j}x_0 + i \epsilon_{ij}^k x_k \right) \ , \\
        x_0 \star x_i &= x_0 x_i + \frac{\lambda}{2} x_i \ , \\
        x_0 \star x_0 &= x_0 \left( x_0 + \frac{\lambda}{2} \right) = \sum_i x_i \star x_i - \lambda x_0 \ ,        
    \end{split}
\end{equation}
resulting in the star-commutation relation
\begin{equation}
    [x_i, x_j]_\star = i \lambda \epsilon_{i j}^k x_k \ . \label{LNC}
\end{equation}

Hence, we have successfully implemented  the linear noncommutativity we were seeking. It is important to emphasize that Eq.~\eqn{LNCsp} is a new $\star$-product by definition.
Moreover, notice that $x_0$ star-commutes with $x_i$, which is also evident from the fact that
\begin{equation}
    \sum_i x_i^2 = x_0^2 \ .
\end{equation}
Therefore, $\mathbb{R}^3_\lambda=(\mathcal{F}(\R^3), \star_\lambda )$ can also be defined as the star-commutant of $x_0$ with respect to the star-product in Eq.~\eqn{LNCsp}. 
The commutative limit is recovered when $\lambda \to 0$. 
According to \cite{Gracia-Bondia:2001ynb}, the same procedure can be applied to get any other type of linear noncommutativity in three dimensions. For another example, we refer the reader to \cite{Pachol:2015qia}, where a star-product for  the \textit{$\kappa$-Minkowski} spacetime is obtained using the same technique, by replacing the $\mathfrak{su}(2)$ generators with those of the $\mathfrak{sl}(2, \C)$ Lie algebra\footnote{This is the Lie algebra of the group of complex $2\times 2$ upper triangular matrices with  unit determinant and real diagonal.}.

\subsection{The matrix basis for the algebra of \texorpdfstring{$\mathbb{R}^3_\lambda$}{}}

In noncommutative field theory, it is convenient to introduce a matrix basis so that noncommutative functions can be represented as (finite or infinite) matrices, making star-products become matrix products. Consequently, the integral in the action functional of noncommutative field theory is replaced by a  trace. For instance,  a matrix basis for the Moyal algebra was first introduced in \cite{Gracia-Bondia:1987ssw, Varilly:1988jk} and was successfully applied in \cite{Grosse:2003nw, Grosse:2004yu} to show the renormalizability to all orders of the $g \varphi^4$ models. Similarly, a matrix basis was introduced for the Wick--Voros product in \cite{Lizzi:2003hz, Lizzi:2003ru, Lizzi:2005zx}. Thanks to the Jordan--Schwinger map described above, this basis could be projected onto a matrix basis for $\mathbb{R}^3_\lambda$ in \cite{Vitale:2012dz}, where it was applied to $\varphi^4$ field theory, and in \cite{Gere:2013uaa}  to gauge models.  For a review on matrix basis for star-products, see \cite{Lizzi:2014pwa}.

According to the literature cited above, matrix basis elements for $\mathbb{R}^3_\lambda$ read
\begin{equation}
    v_{m \Tilde{m}}^j (x) = \frac{e^{- 2 \frac{x_0}{\lambda}}}{\lambda^{2 j}} \frac{(x_0 + x_3)^{j + m} (x_0 - x_3)^{j - \Tilde{m}} (x_1 - i x_2)^{\Tilde{m} - m}}{\sqrt{(j + m)! (j - m)! (j + \Tilde{m})! (j - \Tilde{m})!}} \ ,
\end{equation}
with $j \in \frac{\mathbb{N}}{2}$ and $- j \leq m , \Tilde{m} \leq j$. Any arbitrary function can then be expanded as follows:
\begin{equation}
    \phi (x) = \sum_{j \in \frac{\mathbb{N}}{2}} \sum_{m , \Tilde{m} = - j}^j \phi_{m \Tilde{m}}^j v_{m \Tilde{m}}^j (x) \ ,
\end{equation}
where $\phi_{m \Tilde{m}}^j$ are complex coefficients. Noticing that $v_{m \Tilde{m}}^j (x)$ is a suitable operator symbol of the operator $|j+m, j-m\rangle\langle j+\tilde m, j-\tilde m|$, one can demonstrate that the basis elements are orthogonal, namely
\begin{equation}
    v_{m \Tilde{m}}^j \star v_{n \Tilde{n}}^{\Tilde{j}} (x) = \delta_{\Tilde{m} n} \delta^{j \Tilde{j}} v_{m \Tilde{n}}^j (x) \ .
\end{equation}
Hence, the star-product in $\mathbb{R}^3_\lambda$ transforms into a block-diagonal infinite-matrix product of the form
\begin{equation}
    \phi \star \psi = \sum_{j, m_1, \Tilde{m}_2} \phi_{m_1 \Tilde{m}_1}^j \psi_{m_2 \Tilde{m}_2}^j v_{m_1 \Tilde{m}_1}^j \star v_{m_2 \Tilde{m}_2}^j = \sum_{j, m_1, \Tilde{m}_2} \left(\Phi^j \cdot \Psi^j \right)_{m_1 \Tilde{m}_2} v_{m_1 \Tilde{m}_2}^j \ ,
\end{equation}
where $\Phi$ is a block-diagonal infinite matrix, such that each block is the $(2j + 1) \times (2j + 1)$ matrix $\Phi^j = \{\phi_{mn}^j\}$ with $-j \leq m, n \leq j$.  In particular, one can show that the coordinate functions are represented in the matrix basis as
\begin{equation}
    \begin{split}
        x_- &= \lambda \sum_{j m} \sqrt{(j - m)(j + m + 1)} v_{m m + 1}^j \ , \\
        x_+ &= \lambda \sum_{j m} \sqrt{(j + m)(j - m + 1)} v_{m m - 1}^j \ , \\
        x_3 &= \lambda \sum_{j m} m v_{m m}^j \ , \\
        x_0 &= \lambda \sum_{j m} j v_{m m}^j \,,       
    \end{split}
\end{equation}
where $x_{\pm} = x_1 \pm i x_2$. Therefore, we obtain the following star-product of coordinate functions with basis elements:
\begin{equation}\label{coordsprod}
    \begin{split}
        x_- \star v_{m \Tilde{m}}^j &= \lambda \sqrt{(j + m)(j - m + 1)} v_{m - 1 \Tilde{m}}^j \ , \\
        x_+ \star v_{m \Tilde{m}}^j &= \lambda \sqrt{(j - m)(j + m + 1)} v_{m + 1 \Tilde{m}}^j \ , \\
        x_3 \star v_{m \Tilde{m}}^j &= \lambda m v_{m \Tilde{m}}^j \ , \\
        x_0 \star v_{m \Tilde{m}}^j &= \lambda j v_{m \Tilde{m}}^j \ .
    \end{split}
\end{equation}
Analogous expressions are obtained when considering the coordinate functions acting from the right. We observe that this basis diagonalizes the coordinates $x_3$ and $x_0$, which has eigenvalues $\lambda m$ and $\lambda j$, respectively. In contrast, the coordinates $x_-$ and $x_+$ change the value of $m$ (decreasing and increasing by one, respectively). This also applies with respect to $\Tilde{m}$ when the coordinate functions star-multiply from the right. Note that the value of $j$ remains fixed. This is an important aspect that will be further emphasized later on in this section.

According to the picture given here, the space $\R^3_\lambda$ is the quantum analog of the $\R^3$ foliated into the $S^2$ spheres, with an increasing radius represented by $x_0$. In this spirit, $j$ can be seen as the radius eigenvalue. 

\subsection{Differential calculus over the algebra of \texorpdfstring{$\mathbb{R}^3_\lambda$}{} }

As we discussed in Section \ref{subsec:algebraR3}, the $\mathbb{R}^3_\lambda$ algebra is considered as a subalgebra of the $\mathbb{R}^4_\theta$ algebra when we use the Wick--Voros star-product. 
This identification has an interesting geometric interpretation in the commutative case, where the so-called \textit{Kustaanheimo--Stiefel map} (KS) \cite{Stiefel1965} comes into play.  We will briefly review this setting, which will prove useful for defining a differential calculus and an integral calculus\footnote{We will only review the former. For the latter, which essentially relies on replacing integrals with traces, we refer the reader to  \cite{DAvanzo:2005ohu, Vitale:2012dz, Vitale:2014hca}.}.

The key observation here is the fact that both  $\mathbb{R}^4 - \{0\}$ and $\mathbb{R}^3 - \{0\}$  can be endowed with the structure of trivial bundles over spheres,  being $\mathbb{R}^4 - \{0\} \simeq S^3 \times \mathbb{R}^+$ and $\mathbb{R}^3 - \{0\} \simeq S^2 \times \mathbb{R}^+$. Thus, recognizing $S^3$ as the group manifold of  $SU(2)$, we can use the the so-called \textit{Hopf fibration map} of the principal bundle $SU(2)$ with the base manifold $S^2$ and structure group $U(1)$. This map, denoted as $\pi_H : SU(2) \rightarrow S^2$, allows us to project the derivations of $\mathbb{R}^4 - \{0\}$ down to derivations of $\mathbb{R}^3 - \{0\}$. To clarify, let ${y^\mu}$ be the coordinates in $\mathbb{R}^4$ and ${x^i}$ be the coordinates in $\mathbb{R}^3$. We can view $S^3$ and $S^2$ as submanifolds of $\R^4$ and $\R^3$, respectively, with the constraints $y^\mu y_\mu= 1$  for the former and $x^i x_i= 1$ for the latter. Then, by associating points in  $S^3$  with the $SU(2)$ matrices according to the map
\begin{equation}
    \{y^\mu\}\rightarrow s= y^0 \sigma_0 + i y^i \sigma_i\,,
\end{equation}
where $\det s= 1$, and points in $S^2$ with the $2\times 2$  hermitian matrices
\begin{equation}
    \{x^i\}\rightarrow X=  x^i \sigma_i \,,
    \end{equation}
where $\det X= -1$, the map $\pi_H$ is given by
\begin{equation}
    \pi_H : s \in SU(2) \rightarrow s\sigma_3 s^{\dag}:= x^i \sigma_i\,, 
\end{equation}
such that
\begin{equation}\label{xi}
    x_i= \frac{1}{2}\Tr (\sigma_i s\sigma_3 s^{\dag})\,.
    \end{equation}
The Hopf fibration map is then extended to the principal bundle $\mathbb{R}^4 - \{0\} \rightarrow \mathbb{R}^3 - \{0\}$, with structure group $U(1)$, by relaxing the radial constraints of the $S^3$ and $S^2$ spheres so that $y_{\mu} y^{\mu} = R^2$ and $x^i x_i= r^2$, respectively, where $R, r \in \mathbb{R}^+$. This is the the Kustaanheimo--Stiefel map $\pi_{KS}$, defined as follows:
\begin{equation}
    \pi_{KS} : g \in \mathbb{R}^4 - \{0\} \rightarrow \Vec{x} \in \mathbb{R}^3 - \{0\} \ , \quad  g \sigma_3 g^{\dag} = R^2 s \sigma_3 s^{-1}:= x^k \sigma_k \,, \label{KSm}
\end{equation}
where we have introduced $g = R s$. Analogously to Eq.~\eqn{xi}, the coordinate functions of $\R^3-\{0\}$ are thus obtained as
\begin{equation}\label{xi2}
    x_i= \frac{1}{2}\Tr (\sigma_i g\sigma_3 g^{\dag})\,.
    \end{equation}
It can be easily verified that this expression coincides (up to a factor 2) with the classical Jordan--Schwinger result given in Eq.~\eqref{JSm}, where again we have  $z_1 = \frac{1}{\sqrt{2}} (y_0 + i y_3)$, $z_2 = \frac{1}{\sqrt{2}} (y_1 + i y_2)$, and the identification
\begin{equation}
    x_0 = \frac{R^2}{4} \,.
\end{equation}
where $x_0=r$ is the radius in $\mathbb{R}^3$ and $R$ is the radius in $\mathbb{R}^4$. The KS map was employed in \cite{DAvanzo:2005ohu}  to obtain a basis of derivations for $\mathcal{F} (\mathbb{R}^3 - \{0\})$ by projecting derivations of $\mathcal{F} (\mathbb{R}^4 - \{0\})$. The procedure was later extended in \cite{Vitale:2012dz, Vitale:2014hca} to  the noncommutative setting, where, using the previously mentioned matrix basis,   the restriction to $\mathbb{R}^3 - \{0\}$ can be removed, as the latter is well defined at $0 \in \mathbb{R}^3$.

In the commutative case, a natural basis for derivations in $\mathcal{F} (\mathbb{R}^4)$ is represented by translations $P_{\mu} = \frac{\partial}{\partial y^{\mu}}$. However, since derivations are a left module, we can equivalently choose a basis that is adapted to the respective foliation by means of three-spheres and to the subsequent projection to two-spheres.  
The appropriate basis in this case consists of $\mathfrak{su}(2)$ generators, which are tangent vector fields to $S^3$,
\begin{equation}
    Y_i = y^0 \frac{\partial}{\partial y^i} - y^i \frac{\partial}{\partial y^0} - \epsilon_{ij}^{k} y^j \frac{\partial}{\partial y^k} \ , \quad i= 1\dots, 3\,,
\end{equation}
 and the dilation generators,
 \begin{equation}
     Y_d = y^{\mu} \frac{\partial}{\partial y^{\mu}}\,.
 \end{equation}
Let us then denote $\{D^{(4)}_\mu\} = (\{Y_i\}, Y_d) $ as the set generating the whole left module of derivations for $\mathcal{F} (\mathbb{R}^4-\{0\})$, and hence defining the derivations of $\mathcal{F} (\mathbb{R}^3-\{0\})$. Furthermore, it is worth mentioning that, if these generators are projectable, then $D^{(3)}_i := (\pi_{KS})_* D^{(4)}_\mu$,  where $\pi_*:\mathfrak{X}(\mathbb{R}^4-\{0\})\rightarrow \mathfrak{X}(\mathbb{R}^3-\{0\})$ is the push-forward map \cite{nakahara2018geometry} induced by $\pi$ on vector fields. We observe that $\mathcal{F} (\mathbb{R}^3-\{0\})$ is the kernel of the vector field:
 \begin{equation}
     Y_0 = y^0 \frac{\partial}{\partial y^3} - y^3 \frac{\partial}{\partial y^0} + y^1 \frac{\partial}{\partial y^2} - y^2 \frac{\partial}{\partial y^1}\,,
     \end{equation}
which is the generator of the $U(1)$ fiber\footnote{We have previously seen that the  $\pi_{KS}$ map defined in Eq.~\eqref{KSm} establishes a principal fiber bundle $\mathbb{R}^4 - \{0\} \rightarrow \mathbb{R}^3 - \{0\}$ with structure group $U(1)$.Then, $\mathcal{F} (\mathbb{R}^3 - \{0\})$ can be mapped to $\mathcal{F} (\mathbb{R}^4 - \{0\})$ through the pull-back  map $\pi_{KS}^*$ \cite{nakahara2018geometry}, $$\pi_{KS}^* : f \in \mathcal{F} (\mathbb{R}^3 - \{0\}) \rightarrow f \circ \pi_{KS} \in \mathcal{F} (\mathbb{R}^4 - \{0\}) \ ,$$ hence $\mathcal{F} (\mathbb{R}^3 - \{0\})$ can be regarded as the subalgebra of $\mathcal{F} (\mathbb{R}^4 - \{0\})$ of constant functions along the fiber $U(1)$.}. Therefore, projectable vector fields are those that commute with $Y_0$.  Direct calculation reveals that this condition is satisfied by $Y_i$ for $ i=1,\dots,3$ and $Y_d$, resulting in
\begin{equation}
    \pi_{KS*} (Y_i) = X_i = {\epsilon_{ij}}^k x^j \frac{\partial}{\partial x^k} \ , \quad \pi_{KS*} (Y_d) = x^i \frac{\partial}{\partial x^i} \ .
\end{equation}
Therefore $D^{(3)}_i:= X_i$ and $\;D^{(3)}_d:= X_d$ generate the entire module of derivations on $\R^3-\{0\}$. The first three generate rotations in three dimensions, and thus are not independent (indeed $x^i X_i= 0$), while $X_d$ is the dilation in $\R^3$. 

In the noncommutative case, it is easily verified that the generators of rotations are star-derivations of the $\mathbb{R}^3_\lambda$ algebra,  since they can   be expressed as inner derivations by means of the star-commutator given by
\begin{equation}
    X_i (\varphi) = - \frac{i}{\lambda} [x_i, \varphi]_\star \ , \quad i = 1, 2, 3\ . \label{LNCd}
\end{equation}
The Leibniz rule is then trivially satisfied, and these generators become independent, unlike the commutative case, due to the presence of the star-product. Indeed, even though there exists a null combination with non-zero coefficients such that  $x^i \star X_i (\phi) + X_i (\phi) \star x^i = 0$, this combination is not a star-derivation, as discussed in Section \ref{derba}, since star-derivations are a module only over the center of the algebra. 
Finally, according to the definition in Section \ref{derba}, these generators are sufficient because only functions belonging to the center of the $\mathbb{R}^3_\lambda$ algebra are annihilated by all of them.

On the other hand, the generators of three-dilations $X_d$  are not derivations anymore since the Leibniz rule is not satisfied. This can be easily checked by applying $X_d$ to the star-product of coordinates. As a consequence, despite having a set of independent derivations for the $\mathbb{R}^3_\lambda$ algebra, (which is technically sufficient), the coordinate $x_0$, associated with the radial direction in $\R^3$, is in the center of the algebra. Consequently, it is annihilated by all available derivations, which are inner. This means that, when defining derivation-based dynamics, we will not be able to appropriately describe the radial dynamics. To address this problem, there is a proposal to enlarge the $\mathbb{R}^3_\lambda$ algebra to have dilations as an outer derivative. We refer to \cite{Vitale:2014hca} for details.

Finally, let us comment on the construction of the Laplacian operator, which will be employed in the next Section \ref{sec:gPhiscalar}.  The problem that arises in this case is addressed in \cite{Vitale:2012dz, Vitale:2014hca} and can be traced back to two main requirements: the sought Laplacian should be built upon derivations of the algebra, and it should reproduce the standard commutative limit when $\lambda$ is set to zero. \footnote{Alternative proposals make direct use of the Moyal structures in $\R^4$ without requiring that they induce derivations in $\R^3_\lambda$ \cite{Galikova2013}.} Therefore, in order to cope with the problem discussed above, a multiplicative operator that is quadratic in $x_0 $ was added in \cite{Vitale:2012dz} to the natural candidate $D_i D^i$. This solution can be interpreted as a way to generate radial dynamics, although the commutative limit does not yield the standard Laplacian in $\R^3$. 

\subsection{The \texorpdfstring{$g \varphi^{\star 4}$ scalar field theory on $\R^3_\lambda$}{}} \label{sec:gPhiscalar}

Once we have the $\mathbb{R}^3_\lambda$ algebra with its star-product and the corresponding derivations (along with a well-defined integration procedure in the matrix basis), the action for the $g \phi^{\star 4}$ scalar field theory becomes well-defined:
\begin{equation}
    S[\phi] = \int \dd^3 x \left(\frac{1}{2} D_i \phi \star D^i \phi - \frac{1}{2} m^2 \phi^{\star 2} - \frac{g}{4!} \phi^{\star 4}\right) \ .
\end{equation}
This is formally the same action as in Eq.~\eqref{nclp4}, with the star-product now given by Eq.~\eqn{LNCsp}. After performing an integration by parts, we obtain
\begin{equation}
    S[\varphi] = \int \dd^3 x \left(- \frac{1}{2} \varphi \star (\Delta + m^2) \varphi - \frac{g}{4!} \varphi^{\star 4}\right) \,,
\end{equation}
where $\Delta$ stands for the Laplacian operator defined as follows \cite{Vitale:2012dz, Vitale:2014hca}:
\begin{equation}\label{lapl}
    \Delta \varphi = \alpha \, D_i D^i \varphi + \beta \, x_0^{\star 2} \star \varphi \,,
\end{equation}
with $\alpha$ and $\beta$ being constant real parameters.  As anticipated, let us remark that the second term contains the desired dilation operator,
\begin{equation}
    x_0 \star \phi = x_0 \phi + \frac{\lambda}{2} x_i \partial^i \phi \,, 
    \label{lapl2}
\end{equation} 
which is indeed necessary to describe radial dynamics in $\R^3$. However, it can be easily checked, using Eqs.~\eqn{coordsprod}, that in the matrix basis $v_{m \Tilde{m}}^j$ the term $x_0^{\star 2}$ is diagonal with eigenvalue $\lambda^2 j^2$, and thus radial dynamics is not established. Indeed, radial dynamics should change the radial eigenvalue $j$, which is clearly not the case here. 

This two-parameter family of models (with parameters $\alpha$ and $\beta$)  has been investigated in the matrix basis at the one-loop level in \cite{Vitale:2012dz}, where it was shown to be free from  UV/IR mixing. Beyond the details of the specific results, these are  interesting toy  models that provide an explicit realization of a coordinate-dependent noncommutativity, allowing for detailed calculations.

\section{Noncommutative gauge theory on Moyal space} \label{DSec2}

Let us go back to the Moyal space $\mathbb{R}^{2n}_\Theta$, namely the noncommutative algebra of functions on $\mathbb{R}^{2n}$ with constant noncommutativity. The aim of this section is to introduce gauge theories on this noncommutative space. In the previous sections, we defined the Moyal algebra and a derivation-based differential calculus to describe the dynamics of the theory. Moreover, in Section \ref{sec:ncgft} we encountered some of the structures that we will need in the following. More specifically, we will define
\begin{itemize}
    \item A noncommutative analog of matter fields, replacing the concept of vector bundles over space(time) as presented in Section \ref{matterfield_commutative}.
    \item A noncommutative analog of gauge connection and curvature.
    \item A group of transformations acting from the left, representing gauge transformations.
\end{itemize}

\subsection{Noncommutative connection, curvature, and gauge transformations}\label{moyalgauge}

The concepts mentioned above will be defined in the following for the Moyal space, but the generalization to other noncommutative spaces is straightforward. We first focus on Maxwell's theory in $\mathbb{R}^{2n}$, which is abelian and has a structure group  $U(1)$.  Consequently, charged matter fields interacting with the electromagnetic field are one-dimensional complex vector fields, namely sections of one-dimensional complex vector bundles over $\mathbb{R}^{2n}$. As discussed in Section \ref{sec:cgft}, 
these can be identified with  one-dimensional complex  modules over $\mathcal{F} (\mathbb{R}^{2n})$. Then, the noncommutative generalization to the Moyal space is  immediate: charged matter fields with $U(1)$ interaction are elements of a  one-dimensional complex right module over $\mathbb{R}^{2n}_\Theta$. We will denote this module by
\begin{equation}
    \mathbb{M} = \mathbb{C} \otimes \mathbb{R}^{2n}_\Theta \ ,
\end{equation}
and endow it with a \textit{Hermitian structure} $h$,  such that\footnote{A Hermitian form $h$ is a (sesquilinear) map $h: \mathbb{M}  \times \mathbb{M}  \rightarrow \mathbb{R}^{2n}_\Theta$ such that the following properties hold:
\begin{equation}
    \begin{split}
     h(\boldsymbol{\psi_1} f, \boldsymbol{\psi_2} g) &= f^{\dagger} h(\boldsymbol{\psi_1}, \boldsymbol{\psi_2}) g \ , \nonumber \\ 
     h(\boldsymbol{\psi_1}, \boldsymbol{\psi_2})^{\dagger} &= h(\boldsymbol{\psi_2}, \boldsymbol{\psi_1}) \ , \nonumber
    \end{split}
\end{equation}
for all $f, g \in \mathbb{R}^{2n}_\Theta$ and $\boldsymbol{\psi_1}, \boldsymbol{\psi_2} \in \mathbb{M} $ \cite{Wallet:2008bq}.}
\begin{equation}
    h(\boldsymbol{\psi_1}, \boldsymbol{\psi_2}) = \psi_1^{\dagger} \star \psi_2 \ .
\end{equation}
The general setup for a non-abelian gauge theory (Yang--Mills theory) in $\mathbb{R}^{2n}$ is similar. The structure group is $SU(N)$, and charged matter fields under $SU(N)$ are $N$-dimensional complex vector fields in the defining/fundamental representation of the group, \textit{i.e.,} sections of $N$-dimensional complex vector bundles over $\mathbb{R}^{2n}$. In Moyal space then, charged matter fields under $SU(N)$ are $N$-dimensional complex right modules over $\mathbb{R}^{2n}_\Theta$, \textit{i.e.}, $\mathbb{M}  = \mathbb{C}^N \otimes \mathbb{R}^{2n}_\Theta$.

We now focus on the Moyal algebra $\mathbb{R}^{2n}_\Theta$, its algebra of derivations denoted as $\text{Der} (\mathbb{R}^{2n}_\Theta)$, and the complex right module $\mathbb{M} $ representing matter fields. According to \cite{Wallet:2008bq}, we will introduce a generalization of the concepts of connection, curvature, and gauge transformations  (see also \cite{Dubois-Violette1988, Dubois-Violette:1994dsm, Dubois-Violette:1988cpg} for more details). A noncommutative connection on $\mathbb{M}$ is defined as the linear map $\nabla: \text{Der} (\mathbb{R}^{2n}_\Theta) \times \mathbb{M}\rightarrow \mathbb{M}$ satisfying the following properties:
\begin{equation}
    \begin{split}
        \nabla_X (\boldsymbol{\psi} f) &= \boldsymbol{\psi} X(f) + \nabla_X (\boldsymbol{\psi}) f\ , \\
        \nabla_{c X} (\boldsymbol{\psi}) &= c \nabla_X (\boldsymbol{\psi}) \ , \\
        \nabla_{X + Y} (\boldsymbol{\psi}) &= \nabla_X (\boldsymbol{\psi}) + \nabla_Y (\boldsymbol{\psi}) \ ,
    \end{split}
\end{equation}
for all $f \in \mathbb{R}^{2n}_\Theta$, $c \in \mathcal{Z} (\mathbb{R}^{2n}_\Theta)$ (the center of the algebra), $X, Y \in \text{Der} (\mathbb{R}^{2n}_\Theta)$, and $\boldsymbol{\psi} \in \mathbb{M}$. Additionally, this connection is considered Hermitian if it satisfies the condition
\begin{equation}
    X h(\boldsymbol{\psi_1}, \boldsymbol{\psi_2}) = h (\nabla_X (\boldsymbol{\psi_1}), \boldsymbol{\psi_2}) + h(\boldsymbol{\psi_1}, \nabla_X \boldsymbol{\psi_2}) \ ,
\end{equation}
for any real derivation $X \in \text{Der} (\mathbb{R}^{2n}_\Theta)$, where $h$ is the previously introduced Hermitian structure. The curvature associated to the noncommutative connection $\nabla$ is defined as the linear map $\boldsymbol{F}(X, Y): \mathbb{M} \rightarrow \mathbb{M}$, which satisfies
\begin{equation}
    \boldsymbol{F}(X, Y) \boldsymbol{\psi} = i \left([\nabla_X, \nabla_Y] - \nabla_{[X, Y]} \right) \boldsymbol{\psi} \ .
\end{equation}
The group of (unitary) gauge transformations of $\mathbb{M}$, denoted as $\mathcal{U} (\mathbb{M})$, is the group of automorphisms of $\mathbb{M}$ verifying compatibility conditions with both the structure of the right $\mathbb{R}^{2n}_\Theta$-module, \textit{i.e.},
\begin{equation}
    g (\boldsymbol{\psi} f) =  g (\boldsymbol{\psi}) f \ ,
\end{equation}
and with the Hermitian structure $h$, \textit{i.e.},
\begin{equation}
    h (g (\boldsymbol{\psi_1}), g (\boldsymbol{\psi_2})) = h(\boldsymbol{\psi_1}, \boldsymbol{\psi_2}) \ ,
\end{equation}
for all $g \in \mathcal{U} (\mathbb{M})$. Finally, let us stress that these definitions can be generalized trivially to any other noncommutative space by simply substituting the Moyal algebra with the correspondent noncommutative algebra.

\subsection{Noncommutative electrodynamics on Moyal space}

Now, let us explore the noncommutative formulation of Maxwell's theory on Moyal space, using the tools introduced above. 
For the $U(1)$ gauge theory,  the basis of $\mathbb{M}$ is one-dimensional, therefore we have $\boldsymbol{\psi} = \boldsymbol{e} \psi \in \mathbb{M}$ with $\psi \in \mathbb{R}^{2n}_\Theta$, $\boldsymbol{e}$ the generator of the module. The (Hermitian) connection is then determined by its action on $\boldsymbol{e}$,
\begin{equation}
    \nabla_X (\boldsymbol{\psi}) = \nabla_X (\boldsymbol{e} \psi) = \boldsymbol{e} X (\psi) + \nabla_X (\boldsymbol{e}) \psi \ ,
\end{equation}
with $\nabla_X (\boldsymbol{e})^{\dagger} = - \nabla_X (\boldsymbol{e})$. The gauge connection one-form $\boldsymbol{A}$ is consequently defined as
\begin{equation}
    \boldsymbol{A}: X \rightarrow \boldsymbol{A} (X) := i \nabla_X (\boldsymbol{e}) \ ,
\end{equation}
for all $X \in \text{Der} (\mathbb{R}^{2n}_\Theta)$. In this case, the algebra of derivations is generated by translations (represented as ordinary derivatives in the algebra of functions) $\partial_\mu$, as seen in Section \ref{ScalarF_on_moyal}. In this basis of derivations, we have then
\begin{equation}
    \nabla_{\partial_\mu}^A (\boldsymbol{e}) \equiv \nabla_\mu (\boldsymbol{e}) := - i \boldsymbol{A} (\partial_\mu) = - i \boldsymbol{e} A (\partial_\mu) = - i \boldsymbol{e} A_{\mu} \ ,
\end{equation}
with $A_\mu^\dagger = A_\mu$. Hence
\begin{equation}
    \nabla_\mu (\boldsymbol{\psi}) = \nabla_\mu (\boldsymbol{e} \psi) = \boldsymbol{e} (\partial_\mu \psi - i A_\mu \star \psi) \ .
\end{equation}
One can also show that the curvature two-form $\boldsymbol{F}$ can be written as
\begin{equation}
    \boldsymbol{F} (\partial_\mu, \partial_\nu) = \boldsymbol{e} F (\partial_\mu, \partial_\nu) = \boldsymbol{e} F_{\mu \nu} = \boldsymbol{e} (\partial_\mu A_\nu - \partial_\nu A_\mu - i [A_\mu, A_\nu]_\star) \ .
\end{equation}
The above expression, obtained for the noncommutative abelian gauge theory, resembles that of the commutative non-abelian gauge theories despite the fact that we are still working with $U(1)$. This can be traced back to having a noncommutative algebra of functions. This implies that the noncommutative $U(1)$ gauge theory has distinct features from its commutative counterpart. For example, the self-interaction terms will now appear due to the presence of the Moyal star-product. 
Moreover, all the structures defined above reduce to their commutative analogs for standard $U(1)$ gauge theory in the commutative limit. 

Concerning gauge transformations, the compatibility condition with the Hermitian structure allows to identify gauge transformations with the group of unitary elements of the Moyal algebra. Note that, by considering the gauge transformation $g$,
\begin{equation}
    g (\boldsymbol{\psi}) = g (\boldsymbol{e} \psi) = g (\boldsymbol{e}) \star \psi \ ,
\end{equation}
with $g (\boldsymbol{e}) := \boldsymbol{e} f_g$, the compatibility condition gives
\begin{equation}
    h (g (\boldsymbol{\psi_1}), g (\boldsymbol{\psi_2})) = h (\boldsymbol{e}, \boldsymbol{e}) (f_g \star \psi_1)^\dagger \star f_g \star \psi_2 = h(\boldsymbol{\psi_1}, \boldsymbol{\psi_2}) \ ,
\end{equation}
implying that
\begin{equation}
    f_g^\dagger \star f_g = 1 \ .
\end{equation}
Then, we can conclude that $f_g \in \mathcal{U} (\mathbb{R}^{2n}_\Theta)$, and in turn gauge transformations, form the group of unitary elements of $\mathbb{R}^{2n}_\Theta$ (acting on $\mathbb{R}^{2n}_\Theta$ from the left). As such, we can write
\begin{equation}
    f_g = e_\star^{i \alpha (x)} \ ,
\end{equation}
where the \textit{star-exponential} is defined as
\begin{equation}
   e_\star^{i \alpha}:= \sum_{n = 0}^\infty \frac{i^n}{n!} \underbrace{\alpha (x) \star \dots \star \alpha (x)}_{\text{n times}} \ ,
\end{equation}
and the gauge parameters $\alpha (x)$ are functions in $\mathbb{R}^{2n}_\Theta$.

Finally, we outline the properties of the noncommutative connection and curvature under gauge transformations. The gauge connection is, by definition, gauge covariant. By defining
\begin{equation}
    (\nabla_\mu^A)^g (\boldsymbol{\psi}) := g (\nabla_\mu^A (g^{-1} \boldsymbol{\psi})) \ ,
\end{equation}
we find that
\begin{equation}
    (\nabla_\mu^A)^g (\boldsymbol{\psi}) = \nabla_\mu^{A^g} (\boldsymbol{\psi}) \ ,
\end{equation}
provided that the connection one-form transforms as
\begin{equation}
    A_\mu^g = f_g \star A_\mu \star f_{g^{-1}} + i f_g \star \partial_{\mu} f_{g^{-1}} \ .
\end{equation}
Consequently, one can verify the covariance of the curvature two-form, \textit{i.e.},
\begin{equation}
    F_{\mu \nu}^g = f_g \star F_{\mu \nu} \star f_{g^{-1}} \ .
\end{equation}
As before, gauge transformations for the noncommutative abelian gauge theory are similar to those of a commutative non-abelian gauge theory, recovering the standard $U(1)$ gauge transformations in the commutative limit. The above-outlined properties are important elements because an action that is invariant under gauge transformations can be constructed from them. Indeed, since $F_{\mu \nu}$ is gauge covariant, the term $F_{\mu \nu} \star F^{\mu \nu}$ is gauge covariant as well, and an invariant action is possible because of the cyclicity of the Moyal star-product under integration. Explicitly, upon introducing a metric, 
\begin{equation}
    S [A_\mu] = - \tfrac{1}{4} \int \!\dd^{2n} x \, F_{\mu \nu} \star F^{\mu \nu} \ , \label{NCQED}
\end{equation}
is a natural candidate for the action of the noncommutative $U(1)$ gauge theory on Moyal space. 

\subsection{Spacetime symmetries of electrodynamics on Moyal space} \label{observersymm}

An important feature of the action \eqn{NCQED} resides in the fact that the latter is not only invariant under gauge transformations, but also under standard observer Poincaré transformations. An \textit{observer (or passive) transformation} is essentially a change of coordinates; therefore, everything gets affected by a transformation of this type: background and physical fields (while keeping the underlying unchanged) \cite{Colladay:1998fq}. From the physics point of view,  a passive transformation is better understood as a change of perspective, with a different observer describing the same physical situation.  In this respect, the noncommutativity matrix $\Theta$ in the Moyal space may be regarded as a background field and thus transforms accordingly. It was shown in \cite{Gracia-Bondia:2006nhx} that the Moyal star-product is covariant under passive Poincaré transformations such that the action \eqn{NCQED} turned out to be invariant.  In the cited paper it was shown that the covariance of the Moyal product is actually larger, as it extends to all linear affine transformations, in particular the entire Weyl group, which includes dilations (for earlier references concerning spacetime symmetries of noncommutative algebras also see  \cite{Carroll:2001ws, Iorio:2001qy, Jackiw:2001jb, Bichl:2001yf, Alvarez-Gaume:2003lup}). 

To be more precise, in the remainder of this section, we will focus on the Weyl group. The Weyl group $\text{W} (1, 3)$ extends the Poincaré group by including dilations (the Poincaré group $\text{ISO} (1, 3) = \text{SO} (1, 3) \ltimes \mathbb{R}^4$ is the semi-direct product of the Lorentz group and the group of translations). The action of a Weyl group element  $\Omega = (L, a)$  on a point $x\in \R^4$ is given by
\begin{equation}
    x \longrightarrow x' = \Omega \cdot x = L x + a \,, 
\end{equation}
where $L$ represents the Lorentz group transformations and dilations, and $a$ represents translations. The group law is then given by
\begin{equation}
    \Omega \,\Omega' = (L\, L', La' + a)\,,
\end{equation}
 with the inverse given by $\Omega^{-1} = (L^{-1}, - L^{-1} a)$.
Moreover, the action on  real functions on $\mathbb{R}^4$ is defined as
\begin{equation}
    f (x) \longrightarrow f' (x) = [\Omega \cdot f] (x) = f \left(\Omega^{-1} \cdot x\right) = f \left(L^{-1} (x - a)\right) \ ,
\end{equation}
such that
\begin{equation}
    \Omega_1 \cdot [\Omega_2 \cdot f] = (\Omega_1 \Omega_2) \cdot f \ .
\end{equation}
To check the covariance of the Moyal star-product $\star_\Theta$ under the Weyl group transformations, one needs to compute $[\Omega \cdot f] \star_\Theta [\Omega \cdot g]$. It has been shown in \cite{Gracia-Bondia:2006nhx} that
\begin{equation}
    [\Omega \cdot f] \star_{\Theta} [\Omega \cdot g] = \Omega \cdot (f \star_{\Omega^{-1} \cdot \Theta} g) \ ,
\end{equation}
or equivalently,
\begin{equation}
    [\Omega \cdot f] \star_{\Omega \cdot \Theta} [\Omega \cdot g] = \Omega \cdot (f \star_{\Theta} g) \ ,
\end{equation}
provided that the Moyal noncommutativity tensor $\Theta$ (the background field) transforms as
\begin{equation}
   \Omega \cdot \Theta = L \Theta L^t \ .
\end{equation}
These results show that the Moyal star-product is covariant under standard observer Weyl transformations.  
Therefore, in \cite{Gracia-Bondia:2006nhx}, an $(x, \Theta)$-space was considered to incorporate the following transformation under the action of the Weyl group:
\begin{equation}
    (x, \Theta) \longrightarrow (x', \Theta') = \Omega \cdot (x, \Theta) = (L x + a, L \Theta L^t) \ .
\end{equation}
The infinitesimal generators of these transformations in the $(x, \Theta)$-space are differential operators,
\begin{equation}
    G^\Theta := - (\epsilon^\mu + C_\nu^\mu x^\nu) \partial_\mu - \frac{1}{2} (C_\rho^\mu \Theta^{\rho \nu} + \Theta^{\mu \rho} C_\rho^\nu) \frac{\partial}{\partial \Theta^{\mu \nu}} \ , \label{GT}
\end{equation}
where $C^\mu_\nu= \lambda \delta^\mu_\nu + \omega^\mu_\nu$, and  $\omega$ is an antisymmetric matrix. Let us notice that the second term on the RHS is the Lie derivative of the contravariant skew-tensor $\Theta= \Theta^{\mu\nu}\del_\mu\wedge\del_\nu$ along the vector field $\varepsilon (x) = C_\nu^\mu x^\nu \partial_\mu$,
\begin{equation}
    \left(\mathcal{L}_\varepsilon \Theta\right)^{\mu \nu} = - (C_\rho^\mu \Theta^{\rho \nu} + \Theta^{\mu \rho} C_\rho^\nu) \ . \label{LDT}
\end{equation}
By comparing this with the standard generators in the $x$-space, given by
\begin{equation}
   G :=  - (\epsilon^\mu + C_\nu^\mu x^\nu) \partial_\mu = - (\epsilon^\mu + \lambda x^\mu + \omega_\nu^\mu x^\nu) \partial_\mu \ , \label{G}
\end{equation}
one can easily identify the infinitesimal generators of translations, Lorentz transformations, and dilations,
\begin{equation}
    P_\mu^\Theta = - \partial_\mu \ , \quad M_{\mu \nu}^\Theta = x_\mu \partial_\nu - x_\nu \partial_\mu + \Theta_\mu^\rho \frac{\partial}{\partial \Theta^{\rho \nu}} - \Theta_\nu^\rho \frac{\partial}{\partial \Theta^{\rho \mu}} \ , \quad D^\Theta = - x^\mu \partial_\mu -\Theta^{\mu \nu} \frac{\partial}{\partial \Theta^{\mu \nu}} \ .
\end{equation}
It is then clear that, as expected, the generators of translations remain the same, whereas the others acquire linear contributions in $\Theta$.

The procedure employed above, \textit{i.e.}, showing global covariance and descending to the infinitesimal generators, ensures that the latter   are  derivations of the Moyal algebra, namely, they  satisfy the Leibniz rule according to 
\begin{equation}
    G^\Theta (f \star_\Theta g) = G^\Theta f \star_\Theta g + f \star_\Theta G^\Theta g \ ,
\end{equation}
as in the commutative setting. Moreover,  the new generators obey the same commutation relations as the generators of the standard Weyl Lie algebra. In particular, these relations are
\begin{equation}
    \begin{aligned}
        [P_\mu^\Theta, P_\nu^\Theta] &= 0 \ ,\\
        [M_{\mu \nu}^\Theta, D^\Theta] &= 0 \ ,        
    \end{aligned}\qquad
    \begin{aligned}
        [P_\mu^\Theta, M_{\nu \rho}^\Theta] &= \eta_{\mu \nu} P_\rho^\Theta - \eta_{\mu \rho} P_\nu^\Theta \ , \qquad ~~ [P_\mu^\Theta, D^\Theta] = - P_\mu^\Theta \ ,\\
        [M_{\mu \nu}^\Theta, M_{\rho \sigma}^\Theta] &= \eta_{\mu \sigma} M_{\nu \rho}^\Theta + \eta_{\nu \rho} M_{\mu \sigma}^\Theta - \eta_{\mu \rho} M_{\nu \sigma}^\Theta - \eta_{\nu \sigma} M_{\mu \rho}^\Theta \ ,
    \end{aligned}
\end{equation}
where $\eta$ stands for the Minkowski metric tensor. Then, we can  conclude that the generators $\{P_\mu^\Theta, M_{\mu \nu}^\Theta, D^\Theta\}$ realize the Weyl Lie algebra in the $(x, \Theta)$-space.

Now, let us return to the classical action of the noncommutative electrodynamics on Moyal space \eqref{NCQED} and check its invariance under observer-dependent Weyl transformations. Since $A_\alpha$ is independent of $\Theta$, the action of the Weyl group on the gauge connection, that is given by
\begin{equation}
    A_\alpha (x) \longrightarrow A_\alpha' (x) = [\Omega \cdot A_\alpha] (x) = \Omega \cdot A_\alpha \left(\Omega^{-1} \cdot x\right) \ ,
\end{equation}
must remain unchanged, as in the commutative case. Therefore, the action of the generators  $\{P_\mu^\Theta, M_{\mu \nu}^\Theta, D^\Theta\}$, is  undeformed, namely
\begin{equation}
    \begin{split}
        P_\mu^\Theta [A_\alpha] &= - \partial_\mu A_\alpha \ , \\
        M_{\mu \nu}^\Theta [A_\alpha] &= (x_\mu \partial_\nu - x_\nu \partial_\mu) A_\alpha + \eta_{\mu \alpha} A_\nu - \eta_{\alpha \nu} A_\mu \ , \\
        D^\Theta [A_\alpha] &= - (1 + x^\mu \partial_\mu) A_\alpha \ .
    \end{split}
\end{equation}
Concerning the field strength, it is sufficient to observe that  $\{P_\mu^\Theta, M_{\mu \nu}^\Theta, D^\Theta\}$ are derivations of the star-product. Hence, their respective actions on $F_{\alpha \beta} = \partial_\alpha A_\beta - \partial_\beta A_\alpha - i [A_\alpha, A_\beta]_{\star_\Theta}$ are functionally the same for all values of $\Theta$, including when $\Theta = 0$ and the action on $F_{\mu\nu}$ is then the usual one
\begin{equation}
    \begin{split}
        P_\mu^\Theta [F_{\alpha \beta}] &= \partial_\mu F_{\alpha \beta} \ , \\
        M_{\mu \nu}^\Theta [F_{\alpha \beta}] &= (x_\mu \partial_\nu - x_\nu \partial_\mu) F_{\alpha \beta} + \eta_{\alpha \mu} F_{\nu \beta} - \eta_{\alpha \nu} F_{\mu \beta} + \eta_{\beta \mu} F_{\nu \alpha} - \eta_{\beta \nu} F_{\mu \alpha} \ , \\
        D^\Theta [F_{\alpha \beta}] &= - (2 + x^\mu \partial_\mu) F_{\alpha \beta} \ .
    \end{split}
\end{equation}
This result could be verified directly using the action on the gauge potential and the explicit expression of the deformed generators. 
As a consequence, Eq.~\eqref{NCQED} is  Weyl-invariant in the same way as in the commutative case. As a final remark, we recall that the classical Maxwell action in four spacetime dimensions is also invariant under special conformal transformations, which are quadratic in the coordinates and whose infinitesimal generators  read
\begin{equation}
    K_\mu = x^2 \partial_\mu - 2 x_\mu x^\nu \partial_\nu \ . \label{SCT}
\end{equation}
It was shown in \cite{Gracia-Bondia:2006nhx} that it is not possible to deform such an operator (or, in general, operators that are more than linear in coordinates) to obtain star-derivations. We refer to the cited literature for details. Therefore, we can conclude  that only linear affine transformations (hence, Weyl transformations) remain symmetries of the Moyal star-product. 

\subsection{Problems of QED on Moyal space}

We conclude this section by mentioning two main issues of noncommutative QED on Moyal space, which have been widely analyzed in the past years, namely, the UV/IR mixing and the Gribov ambiguity.

The UV/IR mixing is a feature of noncommutative gauge theories \cite{Hayakawa:1999yt, Matusis:2000jf} which qualitatively manifests itself in the same way as for the scalar field theory considered in Section \ref{ScalarF_on_moyal}. Therefore, similar proposals have been contemplated, such as the addition of harmonic terms to the classical action. The addition of a `harmonic oscillator' term was first proposed in the case of the scalar field theory. The resulting model was remarkably shown to be renormalizable to all orders. This was accomplished initially in the matrix basis for the Moyal algebra, both in two dimensions \cite{Grosse:2003nw} and in four dimensions \cite{Grosse:2004yu}, and then ultimately achieved without employing the matrix basis \cite{Gurau:2005gd}. This modification of the classical action by a harmonic term has also been proposed for QED \cite{deGoursac:2007fzf, Grosse:2007dm}. Alternatively,   noncommutative QED with different star-products beyond constant noncommutativity has been considered. As an example, we will review the case of noncommutative QED on $\mathbb{R}^3_\lambda$ in Section \ref{DSec4}.

The existence of the so-called Gribov copies in noncommutative Moyal QED has been shown in \cite{ Canfora:2015nsa, Blaschke:2016gxl, Kurkov:2017tyn, Holanda:2021lhk}. In standard gauge theory, Gribov ambiguity is a feature of non-abelian gauge theories, consisting in the fact that they exhibit different field configurations that obey the same gauge-fixing condition, yet they are related by a gauge transformation, meaning that they are on the same gauge orbit. This results in an (infinite) overcounting of field configurations. As first shown by Singer \cite{Singer}, and independently by Narasimhan and Ramadas \cite{Narasimhan}, it can be given a precise mathematical characterization in the language of fiber bundles and it only manifests with non-abelian gauge groups. It is also shown in \cite{ Canfora:2015nsa, Blaschke:2016gxl, Kurkov:2017tyn, Holanda:2021lhk} that abelian noncommutative gauge theory behaves in a manner analogous to non-abelian gauge theory in this respect.

\section{Noncommutative gauge theory on \texorpdfstring{$\mathbb{R}^3_\lambda$}{}} \label{DSec4}

Let us go back to the noncommutative space $\mathbb{R}^3_\lambda$, namely the noncommutative algebra of functions in $\mathbb{R}^3$ with linear noncommutativity of coordinate functions of $\mathfrak{su}(2)$ type. In Section \ref{DSec1}, we described the algebra and the corresponding derivation-based differential calculus that allowed for the analysis  of a scalar field theory with quartic self-interaction. To analyze a gauge theory for this spacetime, we need to adapt the analysis performed  in the previous section in the Moyal setting to this case.   Hence, in this section, we will review the construction of  the $U (1)$ gauge theory on the noncommutative space $\mathbb{R}^3_\lambda$ as performed in \cite{Gere:2013uaa}.

\subsection{Noncommutative electrodynamics on \texorpdfstring{$\mathbb{R}^3_\lambda$}{}}

Repeating the analysis performed in \ref{DSec2} for the Moyal case, matter fields are a one-dimensional complex  module over the algebra $\R^3_\lambda$,   $\mathbb{M} = \mathbb{C} \otimes \mathbb{R}^3_\lambda$, namely they are represented as  $\boldsymbol{\psi} = \boldsymbol{e} \psi \in \mathbb{M}$, with $\psi \in \mathbb{R}^3_\lambda$, and $\boldsymbol{e}$ is the generator of the module \cite{Gere:2013uaa}. The derivations of the algebra are inner, namely the action of any derivation $X \in \text{Der} (\mathbb{R}^3_\lambda)$ on any function $\psi \in \mathbb{R}^3_\lambda$ is given by the star-commutator \eqn{LNCd}. The connection and curvature are derived by applying the  definitions from   Section \ref{moyalgauge}, and  gauge transformations are given by the unitary elements of the algebra $\R^3_\lambda$.   

To reformulate the action functional as a matrix model \cite{Gere:2013uaa}, it is convenient to introduce  a fundamental one-form $\boldsymbol{\eta}$, such that
\begin{equation}
    X_i (\psi) \equiv \dd \psi (X^i) = [\eta (X_i), \psi]_\star \ ,
\end{equation}
with $\eta (X_i) = -\frac{i}{\lambda} \in \mathbb{R}^3_\lambda$. One can verify that \cite{Wallet:2008bq}
\begin{equation}
    \nabla_X^\text{inv} (\boldsymbol{\psi}) :=  X (\boldsymbol{\psi}) - \boldsymbol{\eta} (X) \boldsymbol{\psi} = - \boldsymbol{\psi} \boldsymbol{\eta} (X) \ ,
\end{equation}
defines a gauge {\it invariant} connection, $\nabla^\text{inv}$ (known as the canonical connection),  in accordance with the definitions of Section \ref{moyalgauge}. Moreover, 
\begin{equation}
    \boldsymbol{\mathcal{A}} (X) := \nabla_X (\boldsymbol{e}) - \nabla_X^\text{inv} (\boldsymbol{e}) = - i \boldsymbol{A} (X) + \boldsymbol{\eta} (X) \ ,
\end{equation}
where $\nabla_{X_i}(\boldsymbol{e}) $ is the connection one-form $A_i$, defines  a gauge covariant one-form $\boldsymbol{\mathcal{A}}$, which will be useful in the definition of the gauge action. 
 
The derivations of the algebra $\R^3_\lambda$ given by Eq.~\eqn{LNCd}, rescaled by a factor of $\lambda$, 
\begin{equation}
    \text{Der} (\mathbb{R}^3_\lambda) := \left\lbrace D_i := - \frac{1}{\lambda} X_i = \frac{i}{\lambda^2} [x_i, \cdot]_\star \ , \quad i = 1, 2, 3 \right\rbrace \ ,
\end{equation}
close a Lie algebra:
\begin{equation}
    [D_i, D_j] = - \frac{1}{\lambda} \epsilon_{ij}^k D_k \ ,
\end{equation}
with $D_0$ being central, since  $[D_0, D_i] = 0 \ , \ \forall i$, and $D_0 f = 0 \ , \ \forall f \in \mathbb{R}^3_\lambda$. In this basis of derivations, for the gauge connection one-form $\boldsymbol{A}$ we have
\begin{equation}
    \nabla_{D_i} (\boldsymbol{e}) \equiv \nabla_i (\boldsymbol{e}) := - i \boldsymbol{A} (D_i) = - i \boldsymbol{e} A (D_i) = - i \boldsymbol{e} A_i \ ,
\end{equation}
from which we obtain the covariant derivative
\begin{equation}\label{coder}
    \nabla_i (\boldsymbol{\psi}) = \nabla_i (\boldsymbol{e} \psi) = \boldsymbol{e} (D_i \psi - i A_i \star \psi)  \ .
\end{equation}
For the fundamental  one-form $\boldsymbol{\eta}$, in this basis we have
\begin{equation}
    \boldsymbol{\eta} (D_i) = \boldsymbol{e} \eta (D_i) = \boldsymbol{e} \eta_i = \boldsymbol{e} \frac{i}{\lambda^2} x_i \ ,
\end{equation}
so that the gauge invariant connection $\nabla^\text{inv}$ and the gauge covariant one-form $\boldsymbol{\mathcal{A}}$ read, respectively, 
\begin{equation}
    \nabla_i^\text{inv} (\boldsymbol{\psi}) = - \boldsymbol{e} \frac{i}{\lambda^2} \psi \star x_i \ , 
\end{equation}
and 
\begin{equation}
    \boldsymbol{\mathcal{A}} (D_i) = \boldsymbol{e} \mathcal{A} (D_i) =\boldsymbol{e} \mathcal{A}_i = - i \boldsymbol{e} \left(A_i - \frac{1}{\lambda^2} x_i\right) \ .
\end{equation} 
Therefore, the covariant derivative \eqn{coder} can be equivalently rewritten in terms of the gauge covariant one-form
\begin{equation}
    \nabla_i (\boldsymbol{\psi}) = \boldsymbol{e} \left(\mathcal{A}_i \star \psi - \frac{i}{\lambda^2} \psi \star x_i\right) \ .
\end{equation}
Analogously,  the curvature two-form $\boldsymbol{F}$
\begin{equation}
\boldsymbol{F} (D_i, D_j) = \boldsymbol{e} F (D_j, D_j) = \boldsymbol{e} F_{i j} = \boldsymbol{e} \left(D_i A_j - D_j A_i - i [A_i, A_j]_\star + \frac{1}{\lambda} \epsilon_{ij}^k A_k\right)\ ,
  \end{equation}  
 can be rewritten in terms of $\mathcal{A}$ as
 \begin{equation}
    F= \boldsymbol{e} F_{i j} = \boldsymbol{e}\left([\mathcal{A}_i, \mathcal{A}_j]_\star - \frac{i}{\lambda} \epsilon_{ij}^k \mathcal{A}_k\right) \ . 
\end{equation}
Notice that the latter does not contain the derivations anymore, resulting in an action functional that is a matrix model.

It is argued in \cite{Gere:2013uaa} that the gauge-invariant most general action for noncommutative QED on $\mathbb{R}^3_\lambda$ is indeed a  polynomial in the gauge covariant one-form $\mathcal{A}$,
\begin{equation}
    \begin{split}
        S [\mathcal{A}] = \int \dd^3 x &\Big(\alpha \mathcal{A}_i \star \mathcal{A}_j \star \mathcal{A}^j \star \mathcal{A}^i + \beta \mathcal{A}_i \star \mathcal{A}_j \star \mathcal{A}^i \star \mathcal{A}^j \\
        &+ \gamma \epsilon_{ij}^k \mathcal{A}^i \star \mathcal{A}^j \star \mathcal{A}_k + \delta \mathcal{A}_i \star \mathcal{A}^i \Big) \ ,
    \end{split}
\end{equation}
where the Euclidean metric is implicitly introduced. The (real) parameters $\alpha$ and $\beta$ are dimensionless, while $\gamma$ and $\delta$ have mass dimension $1$. In order for the action to be dimensionless, an overall constant usually denoted as $\frac{1}{g^2}$ with mass dimension $- 1$ is needed in three dimensions, but we omit it for simplicity. Going back to the physical field $A$, the  action becomes \cite{Gere:2013uaa}
\begin{equation}
    S [A] = \int \dd^3 x \left(a F_{i j} \star F^{i j} + b \epsilon_{ij}^k \mathcal{A}^i \star \mathcal{A}^j \star \mathcal{A}_k + c \mathcal{A}_i \star \mathcal{A}^i \right) \ ,
\end{equation}
where $a$, $b$, and $c$ are real parameters. This action has the form of a Yang--Mills action (the analog of Eq.~\eqref{NCQED}) with a Chern--Simons term. This theory has been investigated up to one-loop order in the matrix basis, and no UV/IR mixing was found in the analysis \cite{Gere:2013uaa}.

\section{The twist approach to the description of spacetime symmetries} \label{DSec3}

Up to this point, our focus has been on observer (or passive) transformations underlying the spacetime symmetries of QED on Moyal space. In particular, in Section \ref{DSec2} we showed that the Moyal star-product is covariant under linear affine (specifically Weyl) transformations. This aspect eventually led to the invariance of  noncommutative QED on Moyal spacetime under observer-dependent  Weyl transformations. However, the covariance of the Moyal star-product is not preserved by active Weyl transformations. An \textit{active  (or particle) transformation} is performed in a fixed reference frame, meaning that the background fields remain unchanged while the physical fields are  transformed, generally modifying the physics \cite{Colladay:1998fq}. We recall that the background field in our context is the noncommutativity matrix. From the previous section, it can be deduced that the Moyal star-product can no longer be covariant under active Weyl transformations. Nevertheless, the Moyal star-product remains covariant under a \textit{deformed} version of the latter, defined in terms of a \textit{twist} operator. This deformed version gives rise to a \textit{quantum group} of symmetries.

We will first have to review the concept of \textit{Hopf algebra}, and from it that of \textit{twisted Hopf algebra} in order to study the corresponding covariance of the Moyal star-product. The twist approach to describing spacetime symmetries in the noncommutative field and gauge theory was initially introduced in \cite{Oeckl:2000eg} and further investigated in \cite{Chaichian:2004za}. Since then, a large literature has been produced. Early relevant works in this direction include, for instance, \cite{Gonera:2005tm, Gonera:2005hg, Matlock:2005zn, Lizzi:2006xi, Aschieri:2006ye, Chaichian:2006we, Vassilevich:2006tc, Zahn:2006wt} and the references therein.

\subsection{Hopf algebras, twisted Hopf algebras, and twisted symmetries}

In this section, we will briefly summarize the fundamental concepts of Hopf algebras, with more detailed information available in \cite{Chari:1994pz}.

Essentially, a Hopf algebra is an associative algebra, whether commutative or not, with costructures, which are dual to the algebra structures. These structures define a coalgebra, which is (co)-associative, and it can be (co)-commutative or not. Relevant examples include the algebra of smooth functions on a given manifold, $C^\infty(M)$, and the universal enveloping algebra of a Lie algebra, $U(\mathfrak{g})$. The former is both commutative and cocommutative, whereas the latter is not commutative but is cocommutative. We now give the definition of a Hopf algebra.
\begin{definition} [Hopf algebra]
    A \textit{Hopf algebra} $(\mathbb{H}, \mu, i, \Delta, \epsilon, S)$ over a commutative ring $R$ is an $R$-module $\mathbb{H}$ equipped with the following $R$-module maps:
    \begin{itemize}
        \item[i.] The product (or multiplication) map, $\mu: \mathbb{H} \otimes \mathbb{H} \rightarrow \mathbb{H}$.
        \item[ii.] The unit map, $i: R \rightarrow \mathbb{H}$.
        \item[iii.] The coproduct  (or comultiplication map), $\Delta: \mathbb{H} \rightarrow \mathbb{H} \otimes \mathbb{H}$.
        \item[iv.] The counit map, $\epsilon: \mathbb{H} \rightarrow R$.
        \item[v.] The antipode map, $S: \mathbb{H} \rightarrow \mathbb{H}$.
    \end{itemize} 
    These maps need to satisfy certain conditions, ensuring the compatibility of the unital associative algebra $(\mathbb{H}, \mu, i)$ and the counital coassociative coalgebra $(\mathbb{H}, \Delta, \epsilon)$  (see 
    \cite{Chari:1994pz} for details). Hence, $(\mathbb{H}, \mu, i, \Delta, \epsilon)$ is a bialgebra. The additional antipode map $S$ can be considered as a generalization of the inverse map. 
    \end{definition}
As already mentioned, the algebra of smooth functions $C^\infty (G)$  over a Lie group $G$, and the \textit{universal enveloping algebra $U (\mathfrak{g})$} of the Lie algebra $\mathfrak{g}$ of  $G$ naturally have Hopf algebra structures. 
The latter is particularly relevant for the purposes of this section. Let us recall its definition:

\begin{definition}[Universal enveloping algebra of a Lie algebra]
    Let $\mathfrak{g}$ be a Lie algebra over a field $K$. Consider its tensor algebra $T (\mathfrak{g}) = \oplus_{n\geq 0} \mathfrak{g}^{\otimes n}$ (with $\mathfrak{g}^0 = K$) as a vector space. Let $\mathfrak{h}$ be the ideal\footnote{Recall that an \textit{ideal} $\mathfrak{h}$ of a Lie algebra $\mathfrak{g}$ is a Lie subalgebra such that $[\mathfrak{h}, \mathfrak{g}] \subseteq \mathfrak{h}$.} of $T (\mathfrak{g})$ generated by $X \otimes Y - Y \otimes X - [X, Y]$, with $X, Y \in \mathfrak{g}$. The universal enveloping algebra of $\mathfrak{g}$ is $U (\mathfrak{g}) = T (\mathfrak{g}) / \mathfrak{h}$. 
\end{definition}

The  costructures that characterize $U (\mathfrak{g})$ as a Hopf algebra are defined for  Lie algebra elements and then extended. The coproduct, count, and antipode read, respectively, as follows:
\begin{equation}
    \Delta (X) = X \otimes 1 + 1 \otimes X \ , \quad \epsilon (X) = 0 \ , \quad S (X) = - X \ ,
\end{equation}
with $X \in \mathfrak{g}$. In fact, the coproduct is the only map that we will use in the following discussion. For $C^\infty (G)$, the corresponding costructures are given by
\begin{equation}
    \Delta (g) = g \otimes g \ , \quad \epsilon (g) = 1 \ , \quad S (g) = g^{-1} \ ,
\end{equation}
with $g \in G$. Now, we are ready to formally introduce the concept of twist. 

\begin{definition} [Twist] \label{twist}
    Let $\mathbb{H}$ be a Hopf algebra. A twist is an element $\mathcal{F} \in \mathbb{H} \otimes \mathbb{H}$ that is invertible and satisfies the following properties:
    \begin{itemize}
        \item[i.] $(1 \otimes \mathcal{F}) (\mathrm{id} \otimes \Delta) \mathcal{F} = (\mathcal{F} \otimes 1) (\Delta \otimes \mathrm{id}) \mathcal{F}$;
        \item [ii.] $(\epsilon \otimes \mathrm{id}) \mathcal{F} = (\mathrm{id} \otimes \epsilon) \mathcal{F} = 1 \otimes 1$;
    \end{itemize}
   where $\mathrm{id}$ denotes the identity map on $\mathbb{H}$. The first property is referred to as the \textit{2-cocycle condition}.  The second property is simply a compatibility condition with the counit map.

\end{definition}

The twist $\mathcal{F}$ allows for the definition of a new coproduct denoted by $\Delta_\mathcal{F}$ as follows:
\begin{equation}\label{twistedDelta}
    \Delta_\mathcal{F} (h) = \mathcal{F} \Delta (h) \mathcal{F}^{-1} \ ,
\end{equation}
with $h \in \mathbb{H}$. The latter, in turn, defines a new Hopf algebra, called \textit{twisted Hopf algebra} and denoted by $\mathbb{H}_\mathcal{F}$.
Twisted Hopf algebras are concrete examples of quantum groups, namely noncommutative and noncocommutative  Hopf algebras, with the twist inducing a  deformation that only affects the coproduct map. See, for example, \cite{Chari:1994pz, Majid:1996kd} for more details on quantum groups and related topics.

Let us consider a Hopf algebra $\mathbb{H}$, acting on  an associative algebra $\mathcal{A}$, with $m$ denoting the product map in $\mathcal{A}$
\begin{equation}
    m (a \otimes b) = a b \ .
\end{equation}
For $h\in \mathbb{H}$, the action of $h$ on $ab$ is given by
\begin{equation}\label{Leibniz}
    h \triangleright (a b) = h \triangleright m (a \otimes b) = m \circ \Delta (h) \triangleright (a \otimes b) \ ,
\end{equation}
where $\Delta$ represents the coproduct of $H$. We are interested in the case in which $\mathcal{A}$ is the commutative algebra of smooth functions on a spacetime manifold $M$ (with the pointwise product), and $h$ is an element of some  Lie algebra $\mathfrak{g}$, generating the infinitesimal spacetime symmetries of $M$. The latter act on  $\C^\infty(M)$ as derivations. Therefore, in this case, Eq.~\eqn{Leibniz} becomes the Leibniz rule. The twisted coproduct in Eq.~\eqn{twistedDelta}  is compatible with   a deformed or twisted product in $\mathcal{A}$, denoted by $m_\mathcal{F}$, with respect to which Eq.~\eqn{Leibniz} continues to hold. Indeed, defining the twisted   product as
\begin{equation}
    m_\mathcal{F} (a \otimes b) = m \circ \mathcal{F}^{-1}  (a \otimes b) \ , \label{Tp}
\end{equation}
we have
\begin{equation}
    \begin{split}
        h \triangleright m_\mathcal{F} (a \otimes b) &= h \triangleright m \circ \mathcal{F}^{-1}  (a \otimes b) 
        = m \circ \Delta (h) \mathcal{F}^{-1}  (a \otimes b)\\
        &= m \circ \mathcal{F}^{-1} \Delta_\mathcal{F} (h)  (a \otimes b) 
        = m_\mathcal{F} \circ \Delta_\mathcal{F} (h)  (a \otimes b) \ , \label{Tc}
    \end{split}
\end{equation}
namely that, the twisted  algebra $\mathbb{H}_\mathcal{F}$ is represented on  the new algebra $(\mathcal{A}, m_\mathcal{F})$ by its action through $\Delta_\mathcal{F}$. Similarly, it is possible to define an antipode and a counit, which transform $\mathbb{H}_\mathcal{F}$ into a Hopf algebra. Indeed, the associativity of 
 the product \eqn{Tp} and the coassociativity of the coproduct \eqn{twistedDelta} are ensured by the cocycle condition of the twist.

For the case of the commutative algebra of smooth functions considered above,  Eq.~\eqref{Tp} shows how to deform the pointwise product to a noncommutative one using a twist, with the Leibniz rule twisted through Eq.~\eqn{Tc}.  

However, not all star-products are associated with a twist operator. This is the case, for instance, of the Moyal star-product for $\mathbb{R}^4_\Theta$, but not for the Lie-algebra type star-product of $\mathbb{R}^3_\lambda$ in Eq.~\eqn{LNCsp}. 

The consequences of Eq.~\eqref{Tc} are of great importance as well. It implies that a star-product obtained in terms of a twist is always \textit{twist-covariant}. Accordingly, an action functional that is invariant under some spacetime transformations in the commutative case can always be transformed into a noncommutative action, which is invariant under the corresponding \textit{twisted transformations}   (when a twist operator is available). These transformations should be regarded as active or particle-dependent transformations since they do not affect the noncommutativity matrix \cite{Gracia-Bondia:2006nhx}.

\subsection{The Moyal space revisited}

The goal of this section is to recover the Moyal star-product in terms of a twist and, ultimately, analyze the application of the twist approach to the description of spacetime symmetries of the noncommutative QED action given in Eq.~\eqref{NCQED}. We discussed its invariance under standard observer-dependent (passive) Weyl transformations in Section \ref{DSec2}. It can be easily verified that the action cannot be invariant under active Weyl transformations, but only under the subgroup of invariance of the matrix $\Theta$ \cite{Gracia-Bondia:2006nhx}.  

Let us consider the  group of diffeomorphisms of $\mathbb{R}^4$. The Lie algebra of generators consists of all vector fields on $\mathbb{R}^4$ equipped with the usual Lie bracket of vector fields, denoted by $\mathfrak{d} (\mathbb{R}^4)$. We then consider its universal enveloping algebra $U (\mathfrak{d})$, which is a Hopf algebra. In particular, the coproduct is defined for elements of the Lie algebra $h \in \mathfrak{D} (\mathbb{R}^4)$ by $\Delta (h) = h \otimes 1 + 1 \otimes h$, and extended to all of $U (\mathfrak{d})$ through $\Delta (h h') = \Delta (h) \Delta (h')$. The algebra of smooth functions on $\R^4$ with the pointwise product carries a representation of $U (\mathfrak{d})$. It is well-known that the Moyal product in $\mathcal{F} (\mathbb{R}^4)$ can be realized as a twisted product, with a twist  given by 
\begin{equation}
    \mathcal{F}_\Theta = \exp \left\lbrace- \frac{i}{2} \Theta^{\mu \nu} \partial_\mu \otimes \partial_\nu\right\rbrace \ ,
\end{equation}
in terms of the Moyal noncommutativity tensor. The twist is   an element of $U (\mathfrak{d})\otimes U (\mathfrak{d})$, and its inverse is $\mathcal{F}_\Theta^{-1} = \exp \left\lbrace\frac{i}{2} \Theta^{\mu \nu} \partial_\mu \otimes \partial_\nu\right\rbrace$. One can check that the cocycle condition is satisfied as well. The Moyal star-product, as expressed in Eq.~\eqref{moyprod}, follows then from Eq.~\eqref{Tp} according to
\begin{equation}
    m_{\mathcal{F}_\Theta} (f \otimes g) = m \left(\mathcal{F}_\Theta^{-1} \cdot (f \otimes g)\right) = f \star_\Theta g \ .
\end{equation}
Additionally,  any generator of the Lie algebra of diffeomorphisms may be given a twisted coproduct by means of Eq.~\eqref{Tc} \cite{Aschieri:2005yw}
\begin{equation}
    \Delta_{\mathcal{F}_\Theta} (h) = \mathcal{F}_\Theta \Delta (h) \mathcal{F}_\Theta^{-1} \ ,
\end{equation}
ensuring that the  Moyal product $m_{\mathcal{F}_\Theta}$ is covariant under the action of the entire diffeomorphisms algebra.   Hereafter, we will denote these maps as $\Delta\Theta$ and $m_\Theta$.
 
As the classical Maxwell action is invariant under both passive and active Weyl transformations, the deformed action with the Moyal product will be invariant under twisted Weyl transformations.  The twisted coproduct for the generators reads \cite{Chaichian:2004za, Matlock:2005zn} 
\begin{equation}
    \begin{split}
        \Delta_\Theta (P_\mu) &= P_\mu \otimes 1 + 1 \otimes P_\mu \ , \\
        \Delta_\Theta (M_{\mu \nu}) &= M_{\mu \nu} \otimes 1 + 1 \otimes M_{\mu \nu} \\ & \qquad+ \frac{i}{2} \Theta^{\mu \nu} [P_\mu \otimes (\eta_{\alpha \nu} P_\beta - \eta_{\beta \nu} P_\alpha) + (\eta_{\alpha \mu} P_\beta - \eta_{\beta \mu} P_\alpha) \otimes P_\nu] \ , \\
        \Delta_\Theta (D) &= D \otimes 1 + 1 \otimes D - i \Theta^{\mu \nu} P_\mu \otimes P_\nu \ ,
    \end{split}
\end{equation}
with $P_\mu, M_{\mu\nu}, D$ indicating the generators of translations, Lorentz transformations, and dilations, respectively. Except for translations, which are undeformed, the other symmetries get twisted.  We can conclude that, in the context of Moyal noncommutativity, the twist-deformed Maxwell action is invariant under twisted Weyl transformations (Poincar\'e and dilations).  
These should be understood as being of the particle type\footnote{One can demonstrate \cite{Gracia-Bondia:2006nhx} that, given an infinitesimal diffeomorphism which is polynomial in coordinates $$m_\Theta \left( \Delta_\Theta (x^{\alpha_1}\dots x^{\alpha_N})  (x^\mu \otimes x^\nu - x^\nu \otimes x^\mu) \right) = 0 \ ,$$  $\Theta^{\mu \nu}$ remains unchanged. Then, the twist approach accounts only for particle-dependent transformations, as transformations of $\Theta$ cannot be contemplated within this framework.}. Moreover, Eq.~\eqref{Tc} generically holds for any generator $h$ being an infinitesimal diffeomorphism. Hence, Eq.~\eqref{Tc} can be applied to any generator of the universal enveloping algebra. As the Maxwell action in  $3+1$ dimensions has an additional symmetry,  invariance under special conformal transformations,  one can then compute 
the coproduct for the generators of special conformal transformations given in Eq.~\eqref{SCT}, and thus obtain $\Delta_\Theta (K_\mu)$ \cite{Matlock:2005zn, Lizzi:2006xi}. Therefore, in analogy with the commutative case, the noncommutative Maxwell action on Moyal space not only possesses twisted Weyl invariance but also enjoys a twisted conformal invariance. 
We stress once more that this is a feature of twist-deformed models: due to the covariance of the Moyal product under twisted transformations, all symmetries of the untwisted action are mapped to quantum symmetries of the twisted action. 

To conclude, the difference between observer-dependent, or passive transformations, and particle-dependent, or active transformations, in Moyal spacetime amounts to the difference between covariance and twist-covariance of the star-product. This difference can be summarized by the following pair of relations for any given infinitesimal generator $G$:
\begin{equation}
    G^\Theta m_\Theta = m_\Theta \Delta (G) \ ,
\end{equation}
which accounts for covariance, and
\begin{equation}
    \ G m_\Theta = m_\Theta \Delta_\Theta (G) \ .
\end{equation}
which accounts for twist-covariance. We refer to the cited literature for more details.

\section{The twist approach to noncommutative field theory} \label{DSec5}

A different approach to noncommutative gauge theories has been developed within the twist formalism described in Section \ref{DSec3}, for those noncommutative spacetimes that admit a twist. 
The twist approach to noncommutative field and gauge theory is based on twisting not only the product in the algebra of functions, but also the other relevant bilinear maps, establishing, eventually, a well-defined differential calculus, known as a \textit{twisted differential calculus}. In this section, instead of starting with formal definitions, we will go directly into an example and introduce the concepts of twisted differential calculus required for our the discussion. 

\subsection{Angular noncommutativity: the case of \texorpdfstring{$\lambda$}{}-Minkowski} 

Let us consider another noncommutative spacetime exhibiting linear noncommutativity between coordinates. It is indeed simpler than the linear noncommutative space $\mathbb{R}^3_\lambda$ introduced in Section \ref{DSec1}. Its noncommutativity is such that \cite{Dimitrijevic:2017rnf}
\begin{equation}
    [x^0, x^i]_\star = [x^1, x^2]_\star = 0 \ ,\quad [x^2, x^3]_\star = - i \lambda x^1 \ , \quad [x^1, x^3]_\star = i \lambda x^2 \ , \label{ANC}
\end{equation}
where $x^0$ is the time coordinate, $x^i$ (with $i = 1 , 2, 3$) are the space coordinates, and $\lambda$ is a constant with a length dimension. It is also said to reproduce \textit{angular noncommutativity}, since
\begin{equation}
    [x^3, \rho]_\star = 0 \ , \quad [x^3, \varphi]_\star = i \lambda \ ,
\end{equation}
after expressing them in cylindrical coordinates with $x^1 = \rho \cos \varphi$ and $x^2 = \rho \sin \varphi$. It is then clear that the noncommutativity involves the angular coordinates. This is the so-called  \textit{$\lambda$-Minkowski spacetime}\footnote{Indeed, the Lie algebra \eqn{ANC} can be obtained by the $\mathfrak{su}(2)$ Lie algebra underling $\R^3_\lambda$.}. Another variant is the so-called \textit{$\varrho$-Minkowski} spacetime, where the role of the coordinates $x^0$ and $x^3$ is exchanged. This was actually the case in the original work \cite{Dimitrijevic:2017rnf}. Notice that, from an abstract perspective, both are described by the same Lie algebra, but the conceptual differences are notable from a physical point of view. In the former, the time coordinate stays commutative, whereas in the latter it becomes noncommutative. They are, in turn, a variation of the well-known \textit{$\kappa$-Minkowski spacetime} \cite{Lukierski:1991pn, Lukierski:1992dt}, where coordinate functions satisfy $[x^0, x^i]_\star = i/ {\kappa} x^i$ (although as Lie algebras they are quite different). Noncommutative field and gauge theory on $\kappa$-Minkowski spacetime are extensively investigated in the literature, and we refer the reader for more details to \cite{Dimitrijevic:2003wv, Dimitrijevic:2003pn, Grosse:2005iz, Meljanac:2011cs, Dimitrijevic:2011jg, Poulain:2018mcm}. The $\lambda$- and $\varrho$-Minkowski spacetimes with their symmetries have been recently investigated in \cite{Lizzi:2021dud, Lizzi:2022hcq, Fabiano:2023uhg}. 

The Lie algebra underlying Eq.~\eqref{ANC} is  the Euclidean algebra $\mathfrak{e} (2)$ with a central term. Analogously to the case of $\mathbb{R}^3_\lambda$, a star-product reproducing  Eqs.~\eqref{ANC} for coordinate functions was obtained in terms of a Jordan--Schwinger map in \cite{Gracia-Bondia:2001ynb}. Alternatively, and more relevant for this section, it is also possible to obtain another star-product, with the same fundamental noncommutativity \eqn{ANC} in terms of a twist.
In this case, the twist is an element in  $U(\mathfrak{p}  \otimes U(\mathfrak{p}$, where $\mathfrak{p}$ is the Poincaré Lie algebra.  It reads
\begin{equation}
    \mathcal{F} = \exp \left\lbrace- \frac{i \lambda}{2} [\partial_3 \otimes (x^1 \partial_2 - x^2 \partial_1) - (x^1 \partial_2 - x^2 \partial_1) \otimes \partial_3]\right\rbrace \ ,
\end{equation}
and, in cylindrical coordinates, it is expressed as 
\begin{equation}
    \mathcal{F} = \exp \left\lbrace- \frac{i \lambda}{2} (\partial_3 \otimes \partial_\varphi - \partial_\varphi \otimes \partial_3)\right\rbrace \,.\label{ANCt}
\end{equation}
This is an abelian twist\footnote{This is an instance of a more general expression introduced in \cite{Lukierski:1993hk, Lukierski:2005fc, Meljanac:2017oek}.}, meaning that the generators of the Poincaré group considered to construct the twist commute with each other. As a consequence, the associativity of the resulting star-product is automatically guaranteed, since the cocycle condition in Definition \ref{twist} is directly verified. The explicit expression of the twisted star-product reads then
\begin{equation}
    f \star g = m \left(\mathcal{F}^{-1}  (f \otimes g)\right) =  f g + \frac{i \lambda}{2} (\partial_3 f \partial_\varphi g - \partial_\varphi f \partial_3 g) + \mathcal{O} (\lambda^2) \ . \label{ANCsp}
\end{equation}
The twisted differential calculus \cite{Aschieri:2007sq} for this spacetime was first introduced in \cite{Dimitrijevic:2017rnf}. 

It is particularly convenient for developing a calculus in field theory  to compute  the star-product of two plane waves. In the case of the Moyal star-product, we have that (see Appendix \ref{appen_Tadpolediagram}) $e^{i p \cdot x} \star_\Theta e^{i q \cdot x} = e^{i (p + q) \cdot x + p_\mu \Theta^{\mu \nu} q_\nu}$. One can show \cite{Dimitrijevic:2018blz} that, for the star-product in Eq.~\eqref{ANCsp},
\begin{equation}\label{starplane}
    e^{- i p \cdot x} \star e^{- i q \cdot x} = e^{- i (p +_\star q) \cdot x} \ ,
\end{equation}
where we defined the \textit{star-sum} of four-momenta as
\begin{equation}
    p +_\star q = R (q_3) p + R (- p_3) q \ , \label{ANCss}
\end{equation}
with $R$ being the matrix
\begin{equation}
    R (t) \equiv \left( \begin{array}{cccc}
    1 & 0 & 0 & 0 \\
    0 & \cos\left(\frac{\lambda}{2} t\right) & \sin\left(\frac{\lambda}{2} t\right) & 0 \\
    0 & - \sin\left(\frac{\lambda}{2} t\right) & \cos\left(\frac{\lambda}{2} t\right) & 0 \\
    0 & 0 & 0 & 1
    \end{array} \right) \ . \label{R}
\end{equation}
Notice that this is a rotation matrix in the $p_1 p_2$ plane, with the angle of rotation being proportional to the noncommutativity parameter. As expected, it becomes the identity matrix in both the commutative and low-momentum limit. One can check that the star-sum turns out to be noncommutative but associative. It can also be checked that, for an arbitrary four-momentum $k$, we have $k +_\star (- k) = 0$. One can further show that, for the star-product of three plane waves 
\begin{equation}
    e^{- i p \cdot x} \star e^{- i q \cdot x} \star e^{- i r \cdot x} = e^{- i (p +_\star q +_\star s) \cdot x} \ ,
\end{equation}
where
\begin{equation}
    p +_\star q +_\star r = R (q_3 + r_3) p + R (- p_3 + r_3) q + R(- p_3 - q_3) r \ .
\end{equation}
This can be generalized by induction
\begin{equation}
    e^{- i p^{(1)} \cdot x} \star \cdots \star e^{- i p^{(n)} \cdot x} = e^{- i (p^{(1)} +_\star \cdots +_\star p^{(n)}) \cdot x} \ ,
\end{equation}
with
\begin{equation}
    p^{(1)} +_\star \cdots +_\star p^{(n)} = \sum_{j = 1}^n R \left( - \sum_{1 \leq k < j} p_3^{(k)} + \sum_{j < k \leq n} p_3^{(k)} \right) p^{(j)} \ .
\end{equation}

\subsection{The twisted Poincaré Hopf algebra} 

As we commented in Section \ref{DSec3}, the twist in Eq.~\eqref{ANCt} defines a \textit{twisted Poincaré Hopf algebra}, $U_{\mathcal{F}(\mathfrak{p})}$ which is the original Hopf algebra (the universal enveloping algebra of the Poincaré algebra), but with the coproduct map (and antipode) twisted. $U_{\mathcal{F}(\mathfrak{p})}$ was derived  in \cite{Dimitrijevic:2017rnf}, except for the fact that, for the $\lambda$ case, the role of  $\partial_0$ and $\partial_3$ have to be  exchanged. We report here the explicit expression of the twisted coproduct of the generators of the Poincaré algebra, given by
\begin{equation}
    P_\mu = - i \partial_\mu \ , \quad M_{\mu \nu} = i (x_\mu \partial_\nu - x_\nu \partial_\mu) \ ,
\end{equation}
for translations and Lorentz transformations, respectively, and with Lie brackets
\begin{equation}
    \begin{split}
        [P_\mu, P_\nu] &= 0 \ , \\
        [P_\mu, M_{\nu \rho}] &= i (\eta_{\mu \nu} P_\rho - \eta_{\mu \rho} P_\nu) \ , \\
        [M_{\mu \nu}, M_{\rho \sigma}] &= i (\eta_{\mu \sigma} M_{\nu \rho} + \eta_{\nu \rho} M_{\mu \sigma} - \eta_{\mu \rho} M_{\nu \sigma} - \eta_{\nu \sigma} M_{\mu \rho}) \ ,
    \end{split}
\end{equation}
with $\eta$ being  the Minkowski metric tensor. The twisted coproduct of $P_\mu$ is given by
\begin{equation}
    \begin{split}
        \Delta_\mathcal{F} (P_0) &= P_0 \otimes 1 + 1 \otimes P_0 \ , \\
        \Delta_\mathcal{F} (P_1) &= P_1 \otimes \cos\left(\tfrac{\lambda}{2} P_3\right) + \cos\left(\tfrac{\lambda}{2} P_3\right) \otimes P_1 + P_2 \otimes \sin\left(\tfrac{\lambda}{2} P_3\right) - \sin\left(\tfrac{\lambda}{2} P_3\right) \otimes P_2 \ , \\
        \Delta_\mathcal{F} (P_2) &= P_2 \otimes \cos\left(\tfrac{\lambda}{2} P_3\right) + \cos\left(\tfrac{\lambda}{2} P_3\right) \otimes P_2 - P_1 \otimes \sin\left(\tfrac{\lambda}{2} P_3\right) + \sin\left(\tfrac{\lambda}{2} P_3\right) \otimes P_1 \ , \\
        \Delta_\mathcal{F} (P_3) &= P_3 \otimes 1 + 1 \otimes P_3 \ ,
    \end{split}
\end{equation}
while the twisted coproduct of $M_{\mu \nu}$ is given by
\begin{equation}
    \begin{split}
        \Delta_\mathcal{F} (M_{10}) &= M_{10} \otimes \cos\left(\tfrac{\lambda}{2} P_3\right) + \cos\left(\tfrac{\lambda}{2} P_3\right) \otimes M_{10} + M_{20} \otimes \sin\left(\tfrac{\lambda}{2} P_3\right) - \sin\left(\tfrac{\lambda}{2} P_3\right) \otimes M_{20} \ , \\
        \Delta_\mathcal{F} (M_{20}) &= M_{20} \otimes \cos\left(\tfrac{\lambda}{2} P_3\right) + \cos\left(\tfrac{\lambda}{2} P_3\right) \otimes M_{20} - M_{10} \otimes \sin\left(\tfrac{\lambda}{2} P_3\right) + \sin\left(\tfrac{\lambda}{2} P_3\right) \otimes M_{10} \ , \\
        \Delta_\mathcal{F} (M_{30}) &= M_{30} \otimes 1 + 1 \otimes M_{30} - \tfrac{\lambda}{2} P_0 \otimes M_{12} + \tfrac{\lambda}{2} M_{12} \otimes P_0 \ , \\
        \Delta_\mathcal{F} (M_{12}) &= M_{12} \otimes 1 + 1 \otimes M_{12} \ , \\
        \Delta_\mathcal{F} (M_{31}) &= M_{31} \otimes \cos\left(\tfrac{\lambda}{2} P_3\right) + \cos\left(\tfrac{\lambda}{2} P_3\right) \otimes M_{31} + M_{32} \otimes \sin\left(\tfrac{\lambda}{2} P_3\right) - \sin\left(\tfrac{\lambda}{2} P_3\right) \otimes M_{32} \\
        & \qquad- P_1 \otimes \tfrac{\lambda}{2} M_{12} \cos\left(\tfrac{\lambda}{2} P_3\right) + \tfrac{\lambda}{2} M_{12} \cos\left(\tfrac{\lambda}{2} P_3\right) \otimes P_1 \\
        &\qquad- P_2 \otimes \tfrac{\lambda}{2} M_{12} \sin\left(\tfrac{\lambda}{2} P_3\right) - \tfrac{\lambda}{2} M_{12} \sin\left(\tfrac{\lambda}{2} P_3\right) \otimes P_2 \ , \\
        \Delta_\mathcal{F} (M_{32}) &= M_{32} \otimes \cos\left(\tfrac{\lambda}{2} P_3\right) + \cos\left(\tfrac{\lambda}{2} P_3\right) \otimes M_{32} - M_{31} \otimes \sin\left(\tfrac{\lambda}{2} P_3\right) + \sin\left(\tfrac{\lambda}{2} P_3\right) \otimes M_{31} \\
        &\qquad- P_2 \otimes \tfrac{\lambda}{2} M_{12} \cos\left(\tfrac{\lambda}{2} P_3\right) + \tfrac{\lambda}{2} M_{12} \cos\left(\tfrac{\lambda}{2} P_3\right) \otimes P_2 \\
        &\qquad+ P_1 \otimes \tfrac{\lambda}{2} M_{12} \sin\left(\tfrac{\lambda}{2} P_3\right) + \tfrac{\lambda}{2} M_{12} \sin\left(\tfrac{\lambda}{2} P_3\right) \otimes P_1 \ .
    \end{split}
\end{equation}
Note that the coproducts of $P_0$, $P_3$, and $M_{12}$ (the generator of rotations in the $x_1 x_2$ plane) remain undeformed, while the rest become clearly deformed. It can be demonstrated that the twisted coproduct of the momenta $P_\mu$ is indeed related to the star-sum of momenta of Eq.~\eqref{ANCss}  \cite{Dimitrijevic:2018blz}.

\subsection{The twisted differential calculus} 

In the spirit of twist-deforming all bilinear maps, the twist is applied to define a twist-deformed  differential calculus. As for the twist in Eq.~\eqref{ANCt}, the differential calculus was developed in  \cite{Dimitrijevic:2017rnf} (see also \cite{Aschieri:2007sq} and \cite{Aschieri:2005yw, Aschieri:2005zs} for the general setting). 
According to \cite{Aschieri:2007sq}, for an arbitrary bilinear map $\mu$ in some module space of the noncommutative algebra, we have 
\begin{equation}
    \mu : A \times B \rightarrow C \Longrightarrow \mu_\star = \mu \circ \mathcal{F}^{-1} \ .
\end{equation}
Thus, the wedge product of two  forms of arbitrary degree,  $\omega_1$ and $\omega_2$,  is twist-deformed into the star-wedge product
\begin{equation}
    (\omega_1 \wedge_\star \omega_2) (x) = \mathcal{F}^{-1} (y, z) \omega_1 (y) \wedge \omega_2 (z)\Big|_{x = y = z} \ ,
\end{equation}
which is (graded) noncommutative and associative. The Leibniz rule is satisfied by the standard  exterior derivative, namely
\begin{equation}
    \dd (f \star g) = \dd f \star g + f \star \dd g \ .
\end{equation}
This is because the exterior derivative commutes with the Lie derivatives that enter the twisted star product  of Eq.~\eqref{ANCsp} \cite{Dimitrijevic:2018blz}. The nilpotency property, \textit{i.e.}, $\dd^2 = 0$, is verified as well. The standard exterior derivative can then be utilized as the noncommutative exterior derivative. Additionally, the  integral  is  \textit{cyclic} since the twist \eqref{ANCt} is abelian, namely
\begin{equation}
    \int \omega_1 \wedge_\star \cdots \wedge_\star \omega_p = (- 1)^{d_1 d_2 \cdots d_p} \int \omega_p \wedge_\star \omega_1 \wedge_\star \cdots \wedge_\star \omega_{p - 1} \ ,
\end{equation}
with the sum of the degrees of the forms being $d_1 + d_2 + \dots + d_p = 4$. Indeed, the twisted product enjoys  a stronger property: it can be checked to be \textit{closed}, \textit{i.e.},
\begin{equation}
    \int\! \dd^4 x \, f \star g = \int\! \dd^4 x \, g \star f = \int \!\dd^4 x \, f \cdot g \ .
\end{equation}
The closure  of the integral is not fulfilled in general for  star-products. The simplest example where it is violated is the Wick--Voros star product, but also  $\kappa$-Minkowski star-products obtained by Jordanian twists, which are not even cyclic \cite{Pachol:2015qia}, and the linear star-product of $\R^3_\lambda$ that we have discussed previously, which is cyclic but not closed. For details on the closure property of star products and its relation with the Kontsevich approach, see, for example, \cite{Kupriyanov:2015uxa}. 

\subsection{The \texorpdfstring{$g \phi^{\star 4}$ scalar field theory on $\lambda$-Minkowski spacetime} {}  } 

Similarly to the case of $\R^3_\lambda$ discussed previously, the $\lambda$-Minkowski  spacetime with its twisted star-product  and  twisted differential calculus have been considered to study the $g \phi^{\star 4}$ scalar field theory \cite{Dimitrijevic:2018blz}. The action functional  is formally the same as for $\R^3_\lambda$ considered in Section \ref{sec:gPhiscalar} 
\begin{equation}
    S[\varphi] = \int \dd^4 x \left(\frac{1}{2} \partial_\mu \varphi \star \partial^\mu \varphi - \frac{1}{2} m^2 \varphi^{\star 2} - \frac{g}{4!} \varphi^{\star 4}\right) \ . 
\end{equation}
However, the closure property of the star-product under integration allows the removal of the star-product from the quadratic terms of the action. Accordingly, the free (tree-level) propagator of the theory is the same as in the commutative setting. The vertex, as well as the interacting (one-loop) propagator, will change with respect to the commutative counterparts. 

Upon expanding the scalar field $\varphi (x)$ in  Fourier modes,
\begin{equation}
    \varphi (x) = \frac{1}{(2 \pi)^2} \int d^4 p \, e^{- i p \cdot x} \Tilde{\varphi} (p) \ 
\end{equation}
the action is rewritten in momentum space as
\begin{equation}
    \begin{split}
       S[\varphi] &= - \frac{1}{2} \int \dd^4 p \, \dd^4 q \left(p \cdot q + m^2\right) \Tilde{\varphi} (p) \Tilde{\varphi} (q) \delta^{(4)} (p +_\star q) \\
       &- \frac{1}{(2 \pi)^4} \frac{g}{4!} \int \dd^4 p \, \dd^4 q \, \dd^4 r \, \dd^4 s \, \Tilde{\varphi} (p) \Tilde{\varphi} (q) \Tilde{\varphi} (r) \Tilde{\varphi} (s) \delta^{(4)} (p +_\star q +_\star r +_\star s) \ ,
    \end{split}
\end{equation}
where the star-multiplication of plane waves \eqn{starplane} is used. This expression differs from that of the commutative theory due to the appearance in the Dirac delta functions of the star-sum of momenta, which modifies of the momentum conservation in the vertex.   The noncommutative theory is therefore characterized by a \textit{deformed conservation of momentum}. When only two momenta are involved,  this `twisted' conservation of momentum reverts to the usual one. This is because the deformation, encoded in the matrix in Eq.~\eqref{R}, does not affect the conservation of the $p_0$ and $p_3$ components of the momenta, as this matrix represents a rotation in the $p_1 p_2$ plane. Then, we have
\begin{equation}
    \delta^{(4)} (p +_\star q) = \delta^{(4)} \left(R(q_3) p + R(- p_3) q\right) = \delta^{(4)} \left(R(- p_3) (p + q)\right) = \delta^{(4)} (p + q) \ ,
\end{equation}
where we used the property of the $n$-dimensional Dirac delta function, yielding 
\begin{equation}
    \delta^{(n)} (M f) = \frac{1}{|\det M|} \delta^{(n)} (f) \ ,
\end{equation}
for any $n$-component (vector-valued) function $f$ and $n \times n$ non-degenerate matrix $M$. 
his model has been analyzed at one-loop level in momentum space for both the $\lambda$-Minkowski and the $\varrho$-Minkowski spacetimes, revealing UV/IR mixing \cite{Dimitrijevic:2018blz}. As in the Moyal space (see Section \ref{subsec:scalar_moyal}), the one-loop corrections to the propagator result in IR divergences for the non-planar diagram (see Figure \ref{Fig:1loops}), with the planar one being qualitatively the same as in the commutative case.

\section{Concluding remarks}

We have reviewed the mathematical framework of noncommutative gauge and field theory within two main approaches: the derivation-based differential calculus and the twist approach. These were applied to simple underlying geometries, namely the Moyal space and linear noncommutativity.  
Besides the details of the calculations, which necessarily depend on the specifics of the models,  we hope to have conveyed  the main  message of these lecture notes, namely, the importance of the  internal consistency of the construction, which, in the absence of  strong phenomenological indications, should be a guiding tool towards more realistic quantum geometries of spacetime.

In this vein, many  interesting directions of research have been recently considered, such as the $L_\infty$ bootstrap approach \cite{Blumenhagen:2018shf}, which, starting from a given noncommutativity, consistently builds the dynamics and the symmetries of the theory through a sort of   inductive algebraic method. A partially-related approach is Poisson gauge theory \cite{Kupriyanov:2021aet}, which considers a semiclassical limit of noncommutative spacetime, namely a Poisson manifold, and looks for a geometrically consistent  definition of gauge fields and gauge transformations that agree with  the standard gauge theory in the commutative limit. Another promising line of research explores   matrix models \cite{Steinacker:2010rh} as a source of emergent geometry, matter, gauge fields, and gravity. 

Finally, it is important to mention the considerable effort being made to find indirect signatures of the noncommutative nature of spacetime at the energy scales currently accessible. The COST action which hosts these notes is specifically designed for such a purpose. For a recent review, we  refer the reader to  \cite{Addazi:2021xuf}  and references therein.

\section*{Acknowledgements}
The authors acknowledge the contribution of the COST Action CA18108 ``Quantum gravity phenomenology in the multi-messenger approach''. The work of M.~A.~and D.~F.~S.~has been partially supported by Agencia Estatal de Investigaci\'on (Spain) under grant PID2019-106802GB-I00/AEI/10.13039/501100011033, by the Regional Government of Castilla y Le\'on (Junta de Castilla y Le\'on, Spain), and by the Spanish Ministry of Science and Innovation MICIN and the European Union NextGenerationEU (PRTR C17.I1). 
P.~V.~acknowledges support by INFN Iniziativa Specifica GeoSymQFT  and  by the Programme STAR Plus, financially supported by UniNA and Compagnia di San Paolo.

\newpage
\appendix
\section*{Appendices}

\section{The tadpole diagram in \texorpdfstring{$\lambda\varphi^{\star 4}$ on Moyal space}{}}\label{appendix}\label{appen_Tadpolediagram}

We derive the expression of the non-planar tadpole for the $\lambda\varphi^{\star 4}$ theory at one loop and show the correction that the propagator picks up due to the noncommutativity of the product. As we will see and already mentioned in Section \ref{ScalarF_on_moyal}, this non-planar diagram has a different behavior from the planar one. To this aim, we consider the Euclidean action functional for such a field theory that reads
\begin{equation}
    S[\varphi]=\int\! \dd^4 x \, \left( \frac{1}{2}D^\mu \varphi \star D_\mu \varphi - \frac{1}{2} m^2 \varphi^{\star 2} -\frac{1}{4!}\lambda\varphi^{\star 4} \right)\,,
\end{equation}
where $D_\mu \rightarrow \partial_\mu$. Since the product is closed, the free action is the same as the undeformed theory and the tree-level propagators coincide. But the four-vertex is deformed. In momentum space, this is depicted by Figure \ref{Fig:1loops}. In the following, we review the computation which brings to the following expressions for the propagator (trivial) and the vertex, at one loop:
\begin{align}
\label{propagator_vertex}
\Delta^{(0)}=\frac{1}{p^2+m^2}, \qquad V_{\star}=-i \frac{\lambda}{4 !} \delta^3\left(\sum_{a=1}^4 k_a\right) \prod_{a<b} \exp \left(-\frac{i}{2} \theta^{i j} k_{a i} k_{b j}\right)\,.
\end{align} 
To proceed, it is essential to notice that, since we are working in an Euclidean space, therefore there are no sign issues coming from the metric. 
Furthermore, we will see that, up to some point, the spacetime dimension is  not relevant for the purpose of the computation. It will be the case when it comes to investigating the divergences because then, as in the noncommutative case, the latter depends on the dimension of the space.

We can separate the action into a kinetic term and an interaction term as follows:
\begin{align}
    S_k[\varphi]&=\int_{\mathbb{R}^4}\! \dd^4x\left( D_\mu \varphi \star D^\mu \varphi+m^2 \varphi^{\star 2}\right)\,,\\ 
    S_{int}[\varphi]&=\frac{\lambda}{4 !}  \int_{\mathbb{R}^4}\!\dd^4x \, \varphi\star\varphi\star\varphi\star\varphi\,,
\end{align}
where the star-product is the Moyal one and recall that we have
\begin{align}
\label{star_2functions}
    \varphi\star\psi=\varphi(x)\exp\left[\frac{i}{2}\Theta^{\mu\nu}\frac{\overleftarrow{\partial}}{\partial x^\mu}\frac{\overrightarrow{\partial}}{\partial x^\mu}\right]  \psi(x)\,.
\end{align}
\paragraph{Kinetic term.} Upon integration, this term is  equal to the usual one in the commutative theory. To this end, let us show that the Moyal star product is not only cyclic but also closed. 
In order to compute the integral over \eqref{star_2functions}, it is convenient to work in Fourier space because we can then make use of the star-product of plane waves. This explicitly yields
\begin{align}
\label{intgeral_2functions}
      \int \dd^dx\, \varphi\star\psi=\int \dd^dx\int \frac{\dd p^d}{(2\pi)^d}\frac{\dd q^d}{(2\pi)^d}\Tilde{\varphi}(p)\Tilde{\psi}(q)e^{ipx}\star e^{iqx}\,.
\end{align}
We now compute the star-product of the plane waves in the above integral after expanding the exponential function appearing in \eqref{star_2functions}. We then obtain 
\begin{align}
\nonumber
    e^{ipx}\star e^{iqx}&=e^{i(p+q)x}\left(1+\tfrac{i}{2}\Theta^{\mu\nu}\left(ip_\mu\right)\left(iq_\nu\right)+\left(\tfrac{i}{2}\right)^2\Theta^{\mu\nu}\Theta^{\rho\sigma}\left(ip_\mu\right)\left(iq_\nu\right)\left(ip_\rho\right)\left(iq_\sigma\right)+\cdots\right)\\
    \label{straproduct_planewaves}
    &=e^{i(p+q)x} e^{\frac{i}{2}(\Theta^{ij}p_iq_j)}\,,
\end{align}
where, in the last line, we re-identified the series expansion with the exponential function. Hence,  we end up with a new factor that reads $e^{\frac{i}{2}(\Theta^{ij}p_iq_j)}$. The latter is usually written in compact form by introducing the notation $p\wedge q= \Theta^{ij}p_iq_j$. 
On inserting this result in  the integral, we obtain 
\begin{align}
\label{kineticterm_intergral}
 \int \dd^dx\int \frac{\dd^dp}{(2\pi)^d}\frac{\dd^dq}{(2\pi)^d}\Tilde{\varphi}(p)\Tilde{\psi}(q)e^{i(p+q)x}e^{-\frac{i}{2}p\wedge q}&=\int \frac{\dd^dp}{(2\pi)^d}\frac{\dd^dq}{(2\pi)^d}\Tilde{\varphi}(p)\Tilde{\psi}(q)\delta(p+q)e^{\frac{i}{2}p\wedge q}\nonumber\\
 &=\int \frac{\dd q^d}{(2\pi)^d} \Tilde{\varphi}(-p)\Tilde{\psi}(q)\,,
\end{align}
where $p\wedge (-p)=0$.
Therefore, the free action reduces to the undeformed expression
\begin{equation}
    S_k[\varphi]=\int \!\dd^dx\, \frac{1}{2}\varphi\left(-\Box+m^2\right)\varphi\,,
\end{equation}
yielding to the standard two-point Green function
\begin{equation}
    \begin{aligned}
        \left(\Box+m^2\right)G_0(x-y)& = \delta^d\left(x-y\right)\,,\\
 G_0(x-y)&=\int\! \frac{\dd^d }{(2 \pi)^d} \frac{e^{p(x-y)}}{p^2+m^2}\,,
    \end{aligned}
\end{equation}
which allows us to conclude that the tree-level propagator reads
\begin{align}
    \Delta^{(0)}(p)=\frac{1}{p^2+m^2}\,.
\end{align}
The obtained result for the propagator explicitly depends on the kind of star-product that one works with. For instance, for $\kappa$-Minkowski, this is not the case. The closure property of the star-product  turns out to be useful, especially in perturbation theory, since this translates into perturbing around the standard commutative vacuum.

\paragraph{Interaction term.} The vertex in this theory is more interesting because it gives rise to the phase factor in \eqref{propagator_vertex}. To compute it, we have to consider four star-products. We proceed analogously to the previous case and write the interaction term in Fourier space
\begin{align}
    \int \dd^dx\,\frac{\dd^dk_1}{(2\pi)^d}\,\frac{\dd^dk_2}{(2\pi)^d}\,\frac{\dd^dk_3}{(2\pi)^d}\,\frac{\dd^dk_4}{(2\pi)^d}\,\tilde\varphi(k_1)\tilde\varphi(k_2)\tilde\varphi(k_3)\tilde\varphi(k_4)e^{ik_1x_1}\star e^{ik_2x_2}\star e^{ik_3x_3}\star e^{ik_4x_4}\,,
\end{align}
where we can compute, in a similar manner to \eqref{straproduct_planewaves}, the star-product of the four plane waves
\begin{align}
    e^{ik_1x_1}\star e^{ik_2x_2}\star e^{ik_3x_3}\star e^{ik_4x_4}&=e^{i(k_1+k_2)x} e^{-\frac{i}{2}k_1\wedge k_2}\star e^{i(k_3+k_4)x} e^{-\frac{i}{2}k_3\wedge k_4}\,.
\end{align}
After plugging it back in the interaction integral and integrating over the coordinates, the vertex amplitude reads
\begin{align}
\nonumber
&\int \dd^dx\,\prod_{i=1}^4\frac{\dd^dk_i}{(2\pi)^d}\,\tilde\varphi(k_1)\tilde\varphi(k_2)\tilde\varphi(k_3)\tilde\varphi(k_4)
 e^{-\frac{i}{2}\left((k_1 +k_2)\wedge(k_3+ k_4)\right)}e^{-\frac{i}{2}\left(k_1\wedge k_2+k_3\wedge k_4\right)}e^{\sum_i^4 k_ix}\\
 &\qquad=\int\prod_{a=1}^4\frac{\dd^dk_a}{(2\pi)^d}\,\tilde\varphi(k_a)\prod_{a<b}e^{-\frac{i}{2}k_a\wedge k_b}\delta^d\left(\sum_{a=1}^4k_a\right)\,.
\end{align}
In momentum space, we can write the Feynman rules as follows:
\begin{align}
      \Delta_{p}^{(0)}&=\frac{1}{p^2+m^2}\,,\\
      \label{vertex_amp}
      V_k&=\prod_{a<b}e^{-\frac{i}{2}k_a\wedge k_b} \delta^d\left(\sum_{a=1}^4k_a\right)\,.
\end{align}
Now, we can perform the usual techniques using these rules, in an analogous way as we do in standard field theory to derive one-loop corrections. Notice that the vertex acquires a phase factor in comparison to the commutative case. As a consequence, it makes the vertex not invariant under an arbitrary change of momentum flow. We can trace this back to the fact that integrating more than two functions using the Moyal product (as in the free theory) is not invariant. The property of invariance is present only for a cyclic rotation of the factors.  This will affect the one-loop diagrams.

\paragraph{One-loop tadpole.}  At one loop, in this theory there is only one connected diagram that corrects the propagator, and this is the tadpole: since the vertex is a $\varphi^{\star 4}$, we have to match four legs entering the vertex. In the commutative case, we have a multiplicity of twelve that appears in the propagator correction, which cancels the coupling factor in the action.
We compute the tadpole as follows: we have a propagator entering the vertex and then a propagator exiting it again, as one can see in Figure \ref{Fig:1loops}. In general, we have the following momentum relations entering the computation of the amplitude at one-loop:
\begin{equation}
   k_1\wedge k_2+k_1\wedge k_3+k_1\wedge k_4+k_2\wedge k_3+k_2\wedge k_4 +k_3\wedge k_4\,.
\end{equation}
For the planar contributions, we have then 
\begin{equation}
    k_1=-k_4=p\,,\qquad k_2=-k_3=q\,,
\end{equation}
where the minus sign stands for existing momentum $q$. Hence, in the case of the planar loop correction, we have three propagators entering the amplitude, namely two contributions assigned momenta  $p$ and one associated with $q$. Now, after taking into account the symmetry factor, the integration over the momentum $q$, and the vertex amplitude derived in Eq.~\ref{vertex_amp}, the planar integral reads
\begin{align}
    G_2^p=\frac{\lambda^2}{3}\int\frac{\dd^dq}{(2\pi)^d}\frac{e^{\frac{-i}{2}(p\wedge q+p\wedge (-q)+q\wedge (-p)+(-q)\wedge (-p))}}{(p^2+m^2)^2(q^2+m^2)}\,,
\end{align}
and since we have $k_1=k_4$, the contribution $k_1\wedge k_4=k_2\wedge k_3=0$ gives 
\begin{align}
\label{planar_cancelation}
    p\wedge q+p\wedge (-q)+q\wedge (-p)+(-q)\wedge (-p)=0\,.
\end{align}
Thus, the exponential factor above is identically one, and the planar contribution to the one-loop correction reads
\begin{equation}
    \Delta^{(1)}_p=\frac{\lambda^2}{3}\int\frac{\dd^dq}{(2\pi)^d}\frac{1}{q^2+m^2}\,.
\end{equation}
The planar tadpole is then exactly the same because the vertex will be the standard one.

For non-planar cases, we are faced with a different situation, namely 
\begin{align}
    k_2=-k_4=q\,,\qquad k_1=-k_3=p\,,
\end{align}
and therefore we will not have the above cancellation in Eq.~\ref{planar_cancelation} but rather the following additional term:
\begin{align}
     p\wedge (-q)+p\wedge q+(-q)\wedge (-p) + (-p)\wedge q =-2q\wedge p \,,
\end{align}
that will appear in the exponential. Hence, the non-planar term gives 
\begin{equation}
    \Delta^{(1)}_{np}=\frac{\lambda^2}{6}\int\!\frac{\dd^dq}{(2\pi)^d}\frac{e^{iq\wedge p}}{q^2+m^2}\,.
\end{equation}
It is then clear that the contribution due to $q$ does not cancel as in the planar case, and it is precisely coming from the Moyal phase. 

\paragraph{Divergences.} To understand divergences, one should go to the effective dimension of the underlying spacetime, as we do in standard field theory:
\begin{itemize}
    \item The planar tadpole, for example in four dimensions, is quadratically divergent because we would have two momenta in the numerator and two in the denominator, so for large $q$, it is quadratically divergent, namely it is divergent in the ultraviolet.
    \item In three dimensions, it is linearly divergent, and in two dimensions it is $log$-divergent. Therefore, while doing the computations, one should introduce a cut-off for large momenta.
    \item As for the nonplanar diagram, the oscillating phase softens for any dimension, and in the ultraviolet divergence, it is better behaved for large $q$. But then it gives divergent behavior in small $q$. The calculations can be derived in the case of $d=4$ and it reads
\begin{align}
    \Pi^{(1)}_{np}= \frac{C_1}{(\theta p)^2}+ m^2 C_2 \log (\theta p)^2 + F(p)\,.
\end{align}
It is UV-finite but IR-divergent when inserted in higher loops. It is also worth noticing that the model is non-renormalizable.
\end{itemize}

\section{Properties of gauge connections}

As we saw in Section \ref{sec:ncgft},  matter fields are vector fields, namely sections of a vector bundle (or elements of a  complex right module over the algebra of functions), which can be written in a given basis.  The noncommutative formulation of these fields is detailed in Section \ref{sec:cgft}. In this section, we would like to prove the following properties of the gauge connection: 

\begin{enumerate}
    \item Gauge covariance:
    \begin{equation}
        \left(\nabla_\mu^A\right)^g(\boldsymbol{\psi}):=g\left(\nabla_\mu^A\left(g^{-1}\boldsymbol{\psi}\right)\right)\stackrel{\text { check }}{=}\nabla_\mu^{A^g}(\boldsymbol{\psi})\,,
    \end{equation}
    with
    \begin{equation}
        A_\mu^gf_g \star A_\mu \star f_{g^{-1}}+ f_g \star \partial_\mu f_{g^{-1}}\,.
    \end{equation}
    \item Curvature:
    \begin{equation}
        \begin{aligned}
            \mathrm{F}_{\mu \nu}& :=\left(\left[\nabla_\mu^A, \nabla_\nu^A\right]-\nabla_{\left[\partial_\mu, \partial_\nu\right]}^A\right)\stackrel{\text { check }}{=} 
            \left(\partial_\mu A_\nu-\partial_\nu A_\mu-i\left[A_\mu, A_\nu\right]_{\star}\right)\,, \label{Fmunu}\\
            \mathrm{F}_{\mu \nu}^g &=\left(\left[\nabla_\mu^{A_g}, \nabla_\nu^{A_g}\right]-\nabla_{\left[\partial_\mu, \partial_\nu\right]}^{A_g}\right) \stackrel{\text { check }}{=} 
            \left(f_g \star F_{\mu \nu} \star f_{g^{-1}}\right)\,,
        \end{aligned}
    \end{equation}
    implying
    \begin{equation}
        F_{\mu \nu}^g \star F_{\mu \nu}^g=f_g \star F_{\mu \nu} \star F_{\mu \nu} \star f_{g^{-1}}\,,
    \end{equation}
    with  $f_g\in \mathcal{U}(\mathcal{A})=\{f\in \mathcal{A} \, {\rm s.t.}  f^\dag f= 1\}$.
\end{enumerate}

Let us start with the first property, namely the gauge covariance of the connection. Relying on the given definition of the gauge transformation, and bearing in mind that the gauge transformation acts only on the module generator $\mathrm{e}$, as $ g(\mathrm{e}) := \mathrm{e} f_g $, we can write
\begin{equation}
g(\boldsymbol{\psi})=g(\mathrm{e} \psi)=g(\mathrm{e}) \star \psi=\mathrm{e} f_g \star \psi\,,
\end{equation}
and we end up with
\begin{equation}
g\left(\nabla_\mu^A\left(g^{-1}\boldsymbol{\psi}\right)\right)=g\left(\nabla_\mu^A\left(g^{-1}(\mathrm{e}){\psi}\right)\right)=g\nabla_\mu^A\left(\mathrm{e}f_{g^{-1}}\star{\psi}\right)\,.
\end{equation}
By applying the Leibniz rule and the definition of the connection  one-form, \textit{i.e.}, $A_\mu=\nabla_\mu^A (\mathrm{ e})$, we have
 \begin{align}
g\nabla_\mu^A\left(ef_{g^{-1}}\star\psi\right)&=g\left(\nabla_\mu^A(\mathrm{e})f_{g^{-1}}\star\psi+\mathrm{e}\partial_\mu\left(f_{g^{-1}}\star\psi\right)\right)\nonumber\\
 &=g\left(\mathrm{e} A_\mu\star f_{g^{-1}}\star\psi+\mathrm{e} \partial_\mu f_{g^{-1}}\star\psi+\mathrm{e} f_{g^{-1}}\partial_\mu\psi\right)\nonumber\\
 &=\mathrm{e} \left(f_g\star A_\mu\star f_{g^{-1}}\star\psi+f_g\star\partial_\mu f_{g^{-1}}\star\psi+f_g\star f_{g^{-1}}\partial_\mu\psi\right)\,.
 \end{align}
Then, we obtain 
\begin{equation}\label{code}
g\nabla_\mu^A\left(\mathrm{e} f_{g^{-1}}\star\psi\right)= \mathrm{e} \left[\left(f_g\star A_\mu\star f_{g^{-1}} + f_g\star\partial_\mu f_{g^{-1}}\right)\star\psi+ \partial_\mu\psi\right]\,.
\end{equation}
We can then recognize the gauge-transformed connection as follows:
\begin{equation}
\label{gauge_trafo_connection}
    A_\mu^g = f_g \star A_\mu \star f_{g^{-1}} +  f_g \star \partial_{\mu} f_{g^{-1}} \ .
\end{equation}
Therefore, \eqn{code}  is the covariant derivative acting on the matter field $\psi$ with a gauge-transformed connection satisfying the above form.

To prove the transformation property satisfied by the curvature, let us first prove, starting form the definition, that the equality \eqn{Fmunu} holds. We have
\begin{equation}
    {F}_{\mu\nu} \boldsymbol{\psi} =  \left([\nabla_\mu, \nabla_\nu] - \nabla_{[\partial_\mu, \partial_\nu]} \right) (\mathrm{e}\psi) \,,
\end{equation}
where the last term can be omitted because we are using the coordinate basis $\{\del_\mu\}$, which is commutative. By repeatedly using the definition $\nabla_\mu (\mathrm{e}\psi)= \nabla_\mu(e) \psi+ \mathrm{e} \del_\mu\psi)$,  we compute
\begin{align}
\label{curvature}
    \mathrm{F}_{\mu \nu}(\mathrm{e}\psi)&=\left(\left[\nabla_\mu, \nabla_\nu\right]\right)(\mathrm{e}\psi) 
    =\nabla_\mu(\nabla_\nu (\mathrm{e}\psi))- \nabla_\nu(\nabla_\mu(\mathrm{e}\psi))\nonumber\\
    &=\mathrm{e} \left(\partial_\mu A_\nu-\partial_\nu A_\mu\right)\psi+\mathrm{e}\left(A_\mu\star A_\nu-A_\nu\star A_\mu\right)\psi\nonumber\\
    &=\mathrm{e}\left(\partial_\mu A_\nu-\partial_\nu A_\mu-\left[A_\mu, A_\nu\right]_{\star}\right)\psi\,.
\end{align}
To verify the covariance of the field strength, we start from
 \begin{equation}
\mathrm{F}_{\mu \nu}^g= \partial_\mu A^g_\nu-\partial_\nu A^g_\mu-\left[A^g_\mu, A^g_\nu\right]_{\star}\,,
 \end{equation}
 and substitute the gauge transformed potential, Eq.~\eqref{gauge_trafo_connection}. Then, it is a matter of straightforward algebra to check that all non-homogeneous terms cancel out, and we are left with 
 \begin{align}
    F_{\mu \nu}^g &= f_g \star \left( \partial_\mu A_\nu-\partial_\nu A_\mu-\left[A_\mu, A_\nu\right]_{\star}\right) \star f_{g^{-1}}= f_g \star  F_{\mu \nu} \star f_{g^{-1}} \,, 
\end{align}
which is what we wanted to prove.

\addcontentsline{toc}{section}{References}\bibliographystyle{utphys}
\bibliography{bib}

\end{document}